\documentstyle[psfig, 11pt, titlepage]{article}
\renewcommand\baselinestretch{1.0}

\begin{document}

\newcommand{\s}[1]{{\scriptstyle #1}}
\newcommand{\paraf}[2]{\frac{\partial #1}{\partial #2}}
\newcommand{\breuk}[2]{{\scriptstyle \frac{#1}{#2}}}

\baselineskip=16pt
\oddsidemargin=-0.1cm
\evensidemargin=-0.1cm
\topmargin=-1.4cm

\title{On the mechanism of decadal oscillations in a coarse resolution ocean model}
\author{G. Lenderink\thanks{ {\em Corresponding author address:} Ir. G. Lenderink,
        Royal Netherlands Meteorological Institute, P.O. Box 201, 3730 AE De Bilt,
        The Netherlands, e-mail=lenderin@knmi.nl}
        \\ {\em KNMI, De Bilt, The Netherlands} \\ and \\
        R.J. Haarsma \\ {\em KNMI, De Bilt, The Netherlands}  }
\maketitle

\begin{abstract}

The mechanism that causes an interdecadal oscillation in a 
coarse resolution sector ocean model forced by mixed boundary 
conditions is studied. The oscillation is characterized by 
large fluctuations in convective activity and 
air/sea heat exchange on a decadal time scale. When the convective
activity is large, a strengthening of the southeastward
surface flow advects more relatively fresh water from the 
northwestern part of the basin into the convective area, 
which reduces the convective activity. 
Similarly, when the convective activity is small, 
the flow of relatively fresh water is weak,
which enables the expansion of the convective area. 
The periodically strengthening and weakening of the southeastward surface flow
at the polar boundary of the basin is revealed by a negative 
polar cell in the meridional overturning. 
The existence of a halocline and an inverse thermocline
with cool and fresh water above a warm and salty subsurface is essential
for the oscillation.
Further, the oscillation critically 
depends on how the ocean circulation, and especially the surface 
circulation,  responds to anomalous convective activity.
Horizontal boundaries turn out to play an important role
in the dynamical response of the ocean circulation. 
That the dynamical reponse is essential to the oscillation is
confirmed with two simple (conceptual) models, and some 
idealized ocean experiments.  Finally, 
the sensitivity of the oscillation
to salt perturbations and the restoring constant for the air/sea 
heatflux is investigated.

\end{abstract}

\section{Introduction}

In order to determine whether the increasing concentration of greenhouse 
gases is affecting our climate one needs to have a clear picture of natural 
climate variability. It has since long been argued that a considerable part 
of natural climate variability on decadal and longer timescales is due to 
variability of the ocean circulation (Bjerknes 1964; Broecker 1985, Kushnir 1993).
In particular the density driven ocean circulation called thermohaline circulation
(THC) is thought to be responsible for a substantial part of the natural 
climate variability on decadal and longer time scales.

The THC is driven by fluxes of heat and freshwater at the oceans surface. 
The freshwater flux is caused by evaporation, precipitation, river run off 
and sea-ice melt or accretion. The net effect of these fluxes on the surface salinity 
is often accounted for by a virtual salinity flux.  
Simple parameterizations of these heat and salt fluxes are often used
to investigate the variability and stability of the THC. 
Often used are the so-called mixed boundary conditions (MBCs), consisting 
of a fixed salt flux and a time varying heat flux which restores the sea 
surface temperature (SST) towards an apparent atmospheric temperature.
Commonly used values of the restoring constant are 50 W m$^{-2}$ K$^{-1}$ 
(Haney 1971) or even higher, which usage actually pins the SST down to the 
prescribed atmospheric temperature.
MBCs reflect that there is a direct feedback
between the SST and the heatflux, whereas a direct feedback between the 
sea surface salinity (SSS) and the freshwater flux does not exist. However, 
relatively large errors in the heatflux may occur as, using MBCs, the SST 
is forced towards a fixed atmospheric temperature, whereas in reality the 
atmosphere closely adapts to the SST in most areas. 
On the other extreme, some researchers use a fixed heatflux.
This however neglects the possible influence of the ocean on 
climate variability as the ocean is now basically forced to transport a 
fixed amount of heat.

With these relatively simple boundary conditions  variations of the THC on 
decadal and longer time scales were obtained. In this study we will concentrate 
on the (inter)decadal time scale.
Weaver and Sarachik (1991) (hereafter WS91) first found decadal variability 
in an ocean model forced by MBCs. Thereafter many others also 
reported oscillations using different boundary conditions and models 
(Chen and Ghil 1995; Greatbatch and Zhang 1995;
Cai 1995; Weisse et al. 1994; Weaver et al. 1993 (hereafter Wea93); 
Myers and Weaver 1994, Moore and Reason 1993; Huang and Chou 1994; 
Yin and Sarachik 1995 (hereafter YS95); Drijfhout et al. 1996). 
Decadal variability was 
also found with the inclusion of a thermodynamical sea ice model 
(Zhang et al. 1995; Yang et Neelin 1994).

Most of the current research concentrates on the development of boundary conditions 
which parameterize the atmospheric feedbacks more realistically, and the 
investigation of the stability and variability of the THC using these boundary conditions
(Rahmstorf 1994; Cai and Godfrey 1995; Chen and Ghil 1995). 
Although we recognize the importance of this we decided not to
pursue this approach in this paper. We feel that a clear picture of the mechanisms
causing decadal variability using simple boundary conditions is still lacking, 
and that such a picture could be helpful in understanding the oceans behavior
when the atmospheric feedbacks are represented more properly.

Using fixed heat and/or salt fluxes several authors 
(Huang and Chou 1994; Greatbatch and Zhang 1995; Cai 1995, Chen and Ghil 1994)
have noticed the similarity between different forms of variability.  
Greatbatch and Zhang (1995) first noticed 
the similarity of their oscillation using fixed heat and salt fluxes with 
an oscillation obtained in a globally coupled model (Delworth et al. 1993).
Although there is not yet a complete and clear picture of the mechanism of 
these oscillations, there is a growing consensus that these oscillations are 
basically caused by the same mechanism. For MBCs however the picture of 
the mechanisms causing interdecadal variability is more confusing.  
Weaver and Sarachik (1991) explain their oscillation by the advection of a 
positive salt anomaly interacting with convection. Yin and Sarachik (1995), 
however, argue that the same oscillation is mainly set by a 
stabilizing flow of relatively fresh water from the polar boundary, 
and the destabilizing influx of relatively warm water below the surface into 
the convective area. In Moore and Reason (1993) a more complicated mechanism 
of their oscillation is sketched where convection combined with a strong 
negative feedback of anomalous advection of fresh water in the surface 
layer and salty water at a larger depth plays a role. 
They argue that a dynamical interaction between the convective activity and
the strength of the surface flow is essential, whereas Yin and Sarachik (1995) are 
ambiguous on this point. Finally, Zang et al. (1995) argue that their oscillation 
is caused by the thermal insulation effect of sea-ice. 

It is tempting to question whether or not the mechanism causing oscillations 
with MBCs are really so different. But before being able to answer this question
one should have a clear conceptual picture of the mechanisms causing oscillations. 
With this paper we aim to establish a clear picture of the mechanism causing the oscillation
in WS91. This oscillation has also been studied in WS93 and in YS95.
Large variations in convective activity and the air/sea heat flux 
are characteristic of this oscillation. The oscillation turns out to be mainly driven 
by convection and advection. The impact of convection and advection on the heat and salt 
budgets of the convective area is carefully estimated. Our analysis indicates that the 
oscillation crucially depends on how the surface flow responds to anomalous convective 
activity. The advection by the surface flow of fresh water into the convective area mainly 
determines the convective activity. We show that the period of the oscillation is mainly 
determined by the timescale of this advective feedback.  

In this paper we do not describe a new mechanism. Several processes and 
feedbacks that are potentially important are already mentioned in literature. 
Our aim is to clarify the role of these feedbacks and processes, and 
to complete the picture of the mechanism as far as possible.
Yin and Sarachik (1995) already gave a quite extensive description similar to ours of
many important processes. Our conclusions however differ from theirs mainly
concerning the question what determines the time scale and the amplitude of 
the oscillation, and we will therefore focus on these differences.
Our description of the response of the surface flow to anomalous convective 
activity also shows a large resemblance with the  
mechanism described by Moore and Reason (1993). Finally, several authors 
(Winton 1996a; Greatbatch and Peterson 1995) also proposed that boundary waves
might play a role. 
This aspect of the oscillation is also studied in this paper.
  
Besides the question concerning the mechanism of the oscillation, information
about the sensitivity and robustness of the oscillation (to changes in parameter values, 
forcing, inclusion of
other processes like sea-ice and/or atmospheric feedbacks, atmospheric noise or perturbations) 
is important. Two of our experiments shed some light on this.
We show that the oscillation is fairly insensitive to perturbations
in the surface salinity. This contrasts highly with the sensitivity of steady states obtained 
by a spin-up with restoring conditions. With MBCs these steady states usually show 
a high sensitivity to surface salinity perturbations, and a polar halocline catastrophy is 
easily invoked.
Furthermore, we show that the oscillation is rather insensitive to the restoring 
constant for the surface temperature. Our results on this point agree with those in YS95.

The contents of the paper is as follows. In section \ref{h3.sec.weaver} we reproduce 
the oscillation of WS91, and in section \ref{h3.wea.sec.mech} a detailed analysis 
of the oscillation is given. 
In section \ref{h3.sec.sensitive} several sensitivity runs are performed.
A comparision with the results of Yin and Sarachik (1995)
 is made in section \ref{h3.sec.comYS95}. 
A summary and conclusions are given in section \ref{h3.sec.conclusions}.
In the appendix we investigate the possibility of convective oscillations
in a two box model.


\section{The model}

The ocean model used is basically a "GFDL" type model. It uses the usual budget equations 
for temperature, and salinity with diffusion parameterizing the eddy-transports.
The momentum equations are split in a barotropic and a baroclinic part. For the barotropic 
part the "Sverdrup-balance" is solved (see Lenderink and Haarsma 1994, hereafter LH94). 
As we use a flat bottomed ocean the barotropic 
part is completely determined by the windstress only.  
The equations for the baroclinic velocity field are solved in time. 
An asynchronous time integration  (Bryan 1984) is used to slow down internal gravity and Kelvin waves.
In Wea93 and YS95 it has been shown that the asynchronous time integration does not 
significantly affect the oscillation.  
An eddy viscosity is used in both horizontal as vertical direction, whereas the 
horizontal and vertical advection of momentum is neglected.
The equation of state is similar to the one employed in LH94. 
A convective mixing scheme is employed after each timestep. It is the 
standard mixing scheme as employed in the GFDL Bryan-Cox code. 
Starting from the surface level the densities of each
pair of levels (first level 1 and 2, then level 3 and 4 etc.) are compared and the levels are mixed 
completely in case of an unstable stratification. Thereafter, the procedure is 
repeated starting from the second level. 

The model basin extends from the equator to 64 $^o$N and
extends 60 degrees in longitudinal direction. The ocean basin
is 4000 meters deep and bottom topography is not included.
The horizontal resolution is 3.75 $^o$ in longitudinal and 4.00 $^o$ 
in latitudinal direction. 
The layer thicknesses are (from the surface to the bottom) 40, 40, 80, 140,
250, 350, 400, 450, 500, 550, 600 and 600 meters. 
The timestep for the ocean model is 4 days.
For the horizontal and vertical diffusivity we used
$1 \; 10^3$ m$^2$ s$^{-1}$ and $0.6 \; 10^{-3}$ m$^2$ s$^{-1}$ respectively,  and for
the horizontal and vertical viscosity  $2.5 \; 10^5$ m$^2$ s$^{-1}$ and
$1 \; 10^{-3}$ m$^2$ s$^{-1}$. 


\section{The oscillation \label{h3.sec.weaver}}

\subsection{Reproduction of the oscillation}

\begin{figure}[h]
\centerline{\psfig{figure=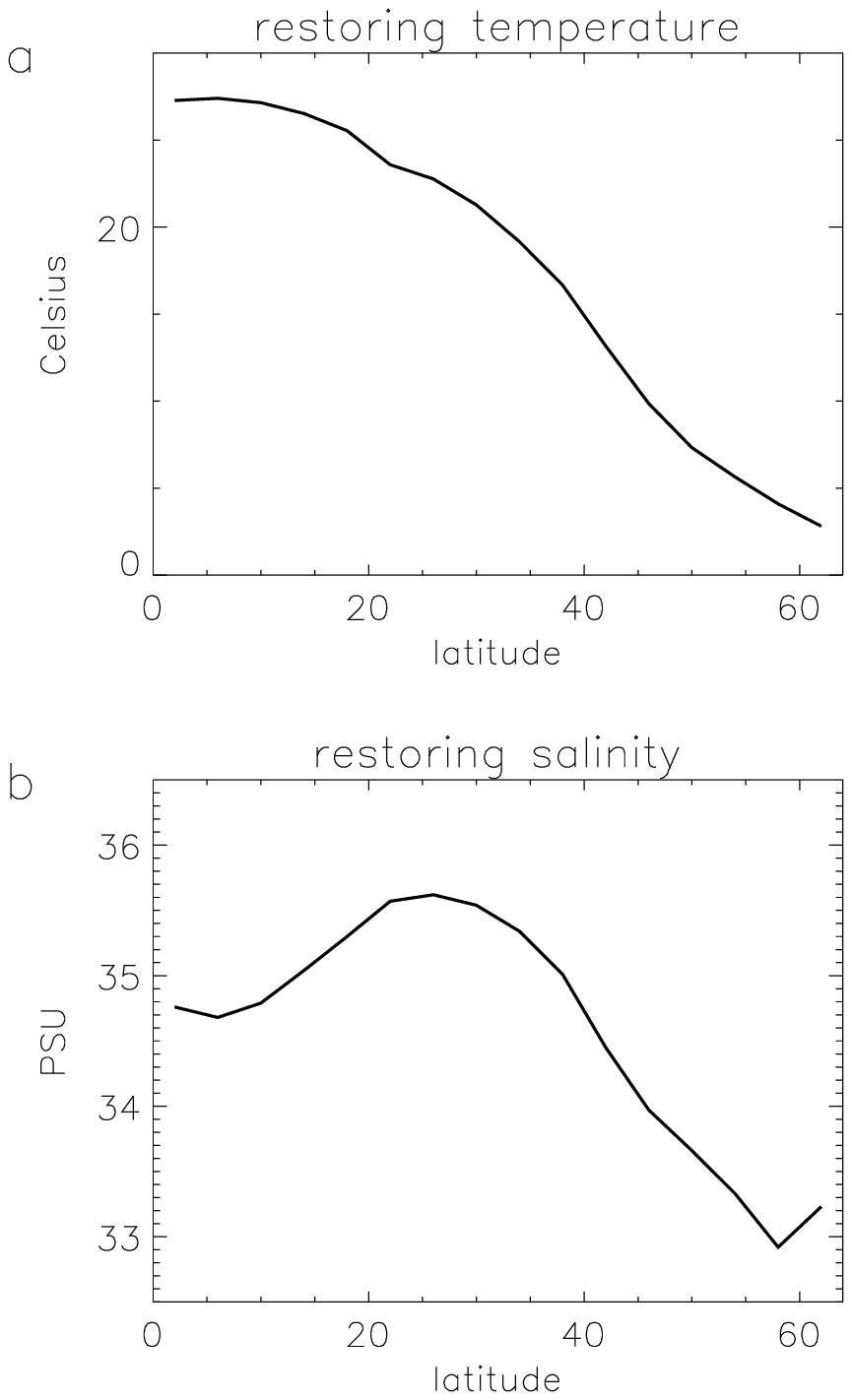,height=7cm,width=7cm,clip=}}
\center{\parbox{14cm}{\renewcommand{\baselinestretch}{1.1}
\caption{\label{h3.wea.forcing} \small
Restoring temperature and salinity for the reproduction experiment.
}}}
\end{figure}

We performed an experiment similar to the experiments described in Wea93 and WS91.
First, we spun our model up from rest using restoring conditions 
We used forcing fields and parameter
values similar to the ones used in Wea93 and WS91. 
The temperature and salinity profile to which the surface is restored, 
which are the same as in WS91, are shown in Fig. \ref{h3.wea.forcing}. 
For the restoring timescale we used 20 days.
The zonal windstress also equals the profile 
used in WS91. 

\begin{figure}[h]
\centerline{\psfig{figure=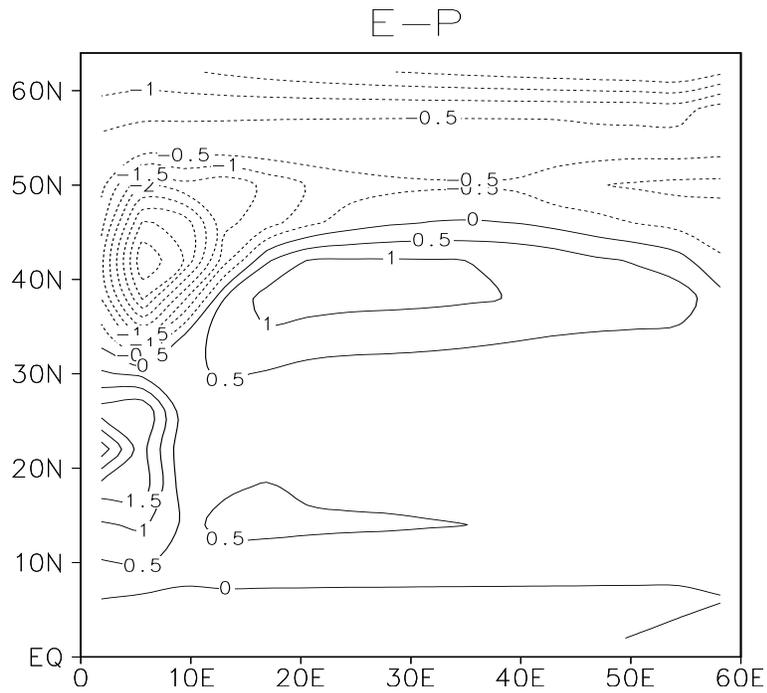,height=9cm,width=10cm,clip=}}
\center{\parbox{14cm}{\renewcommand{\baselinestretch}{1.1}
\caption{\label{h3.wea.emp}
\small 
Diagnosed E-P (evaporation - precipitation) flux in m yr$^{-1}$.
}}}
\end{figure}

\begin{figure}[h]
\centerline{\psfig{figure=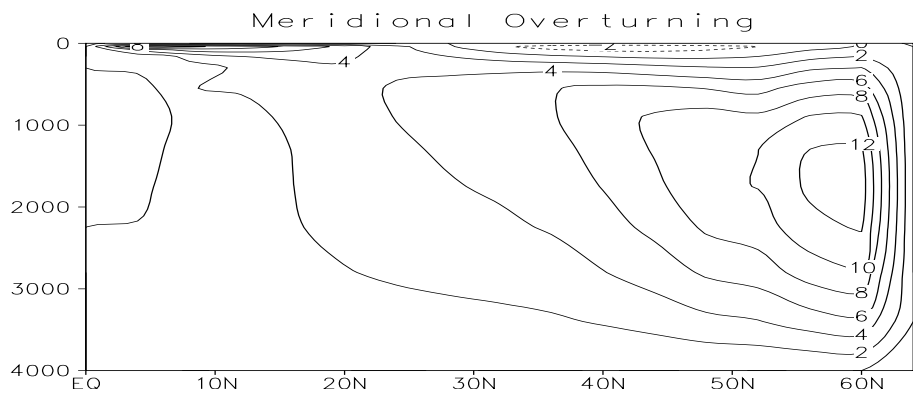,height=5cm,width=12cm,clip=}}
\center{\parbox{14cm}{\renewcommand{\baselinestretch}{1.1}
\caption{\label{h3.wea.overspup}\small 
Meridional overturning (Sv) spin-up circulation as a function of 
latitude and depth.  
}}}
\end{figure}

After the system reached its equilibrium state with restoring conditions, 
we diagnosed the E-P field needed to sustain this state.
This field as shown in Fig \ref{h3.wea.emp} closely resembles Fig. 2c in Wea93.
The spinup state is characterized by deep convection along the 
polar boundary. Relatively shallow convection up to 1000 meter also 
occurs at 50 $^o$N. The meridional overturning circulation as shown 
in Fig \ref{h3.wea.overspup} is very similar to Fig. 6b of Wea93 and consists of a 
positive overturning cell of 13 Sv with sinking motion at the polar boundary. 
Switching to mixed boundary conditions, this state collapses after a small 
salinity perturbation (-0.1 psu in one surface gridbox only) in the convective area.
Deep convective mixing at the polar boundary vanishes, meanwhile, however, 
convective activity at lower latitudes increases in strength and 
the system starts to oscillate on a decadal timescale.
We ran the model another 2500 years during which the system continues to 
oscillate. We analyzed the oscillation after approximately 2000 years of integration.

\begin{figure}[h]
\centerline{\psfig{figure=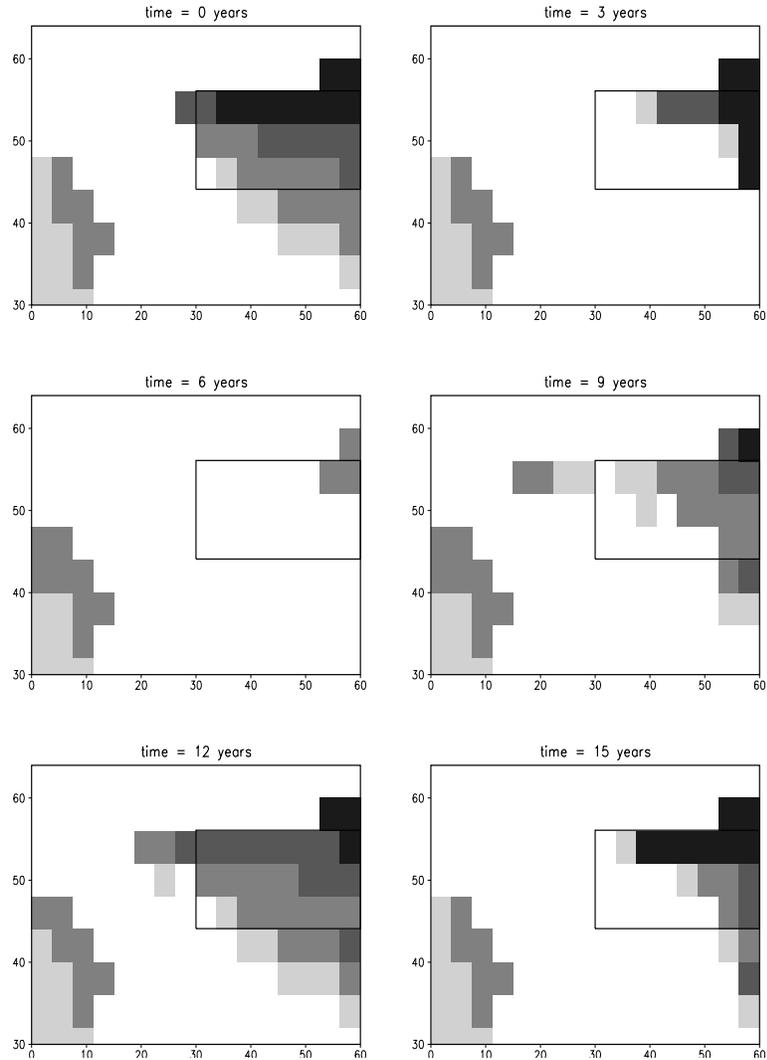,height=14cm,width=10cm,clip=}}
\center{\parbox{14cm}{\renewcommand{\baselinestretch}{1.1}
\caption{\label{h3.wea.con} \small 
Convective activity during oscillation in the polar part of the basin
north of 30 $^o$N. Shading correspond with convective mixing until approximately
(from light to dark) 200 m, 900 m, 2200 m, and 4000 m, respectively.
Also shown is the area where we considered the mean quantities.
}}}
\end{figure}

Fig. \ref{h3.wea.con}  and Fig. \ref{h3.wea.over} show the convective activity 
and the meridional overturning at 6 times during the oscillation. 
We start our description at t = 0 years when the convective activity is strong. 
A relatively weak positive 
overturning cell and a fairly strong negative polar cell are shown. In the next 
3 years shallow convective mixing is decreasing, whereas deep convective
mixing to the ocean bottom continues. At the same time both negative and
positive overturning cell increase in strength. Shortly after t = 4 years deep convective 
mixing disappears completely. The negative overturning cell collapses in
one year time, and the maximum of the overturning displaces to the polar boundary. 
Shallow convection starts again only a few years after the collapse of the negative 
cell, shortly before t = 6 years. In the following years, first a spreading of the 
convective area and a decrease in strength of the positive overturning cell, and then 
a gradual retreat of the convective area, a deepening of convective mixing at the 
northeastern corner and an increase in strength of the overturning cell is observed. 
During these years the polar cell monotonically increases in strength.
After 13 years the circulation approximately equals the circulation at t = 0
and a new cycle starts. 

\subsection{A simple area mean picture}

Fluctuations in convective activity are a dominant feature 
of the oscillation. In response to the convective activity the
atmospheric heatflux and also the oceanic meridional heat transport display 
large variations (see below and also WS91).
We therefore concentrate on the area A$_{con}$ indicated in Fig. \ref{h3.wea.con}. 
The area A$_{con}$ is approximately the smallest rectangular enclosure of 
the area where the convective activity changes during the oscillation.
In Fig. \ref{h3.wea.depth} we show the area mean temperature and salinity as a 
function of depth and time. In this figure we also plotted 
the time evolution of the convective fraction. Here, the convective fraction
is defined as the area affected by convective mixing at a certain depth
divided by the total area A$_{con}$. It is shown that the strongest fluctuations
in temperature and salinity occur in the upper 700 meters of the ocean. 
The main convective activity is confined to the upper 1000 meters. 
In the upper 600 meters about 90 \% of the 
area is affected by convection, and between 600 and 1000 meters still
60 \% of the area is affected. Below 1000 meters, however, this measure 
sharply decreases to about 20 \%. In the following we will therefore concentrate on the upper
part of the ocean. Figure \ref{h3.wea.depth} shows
a warm subsurface and a fresh surface ocean in the nonconvective phase. 
In this phase the salinity gradient (halocline) keeps the ocean stably stratified, 
whereas the temperature gradient (inverse thermocline) tends to destabilize the stratification.
During the convective phase the heat contained by the anomalous warm subsurface 
ocean is rapidly  released to the atmosphere. This heatflux is shown in Fig. \ref{h3.wea.htflux}.
At the same time the surface becomes considerably saltier.

In order to investigate which processes dominate the evolution in the convective  
area A$_{con}$ we defined a surface and a subsurface box in A$_{con}$ and 
investigated their evolution in time. The surface box $B_u$ consists of the uppermost 
three levels (0-160 m) and the subsurface box $B_l$ consists of 
the next two levels (160-550 m). We have chosen these levels for $B_l$  
because in the nonconvective phase these levels contain the 
major part of the heat that is subsequently given off to the atmosphere in the 
convective phase. 

We plotted in Figs. \ref{h3.wea.budget}a,b the mean temperature and salinity
for both boxes. The large variations in the surface (of $B_u$) salinity and
the subsurface (of $B_l$) temperature are again clearly visible. We focus
on the cause of these variations. Thereto, we plotted in 
Figs. \ref{h3.wea.budget}c,d the time derivative of the surface 
salinity and the subsurface temperature.
We also plotted the contribution of horizontal advection and convection.
Figure \ref{h3.wea.budget}c shows that horizontal advection and convection
almost completely determine the surface salinity balance.
Horizontal and vertical diffusion (not shown) are of only of secondary importance
when considering the box averaged values, although locally they could play a role.
The surface heat balance is mainly determined by convection and the atmospheric 
heatflux (not shown). The atmospheric heatflux keeps the surface temperature close to 
the atmospheric temperature.
The fluctuations of the surface box temperature
are mainly caused by the contribution of the temperature of level 3 --
which is less affected by the surface heatflux -- to the box mean.
For the subsurface box $B_l$ horizontal advection provides for a nearly constant source 
of heat and salt. These heat and salt fluxes are mainly due to the meridional 
overturning which advects heat and salt from the equator northward just below the surface.
In the absence of convection the temperature and salinity of the subsurface rise
due to this term. When convective activity is large the convective flux however
causes a decrease in the subsurface temperature and salinity.

A relatively simple picture of the dominant processes in the 
convective area now emerges. First, consider t = 0, when the convective activity is large. 
At this time an increasing amount of fresh water due to horizontal advection enters 
the surface box. The  area with convection is decreasing in response to
the increasing advection of fresh water water into the convective aea.
The surface salinity decreases further because of the decreasing amount of salt mixed up 
by convection. At the same time a gradual warming of the subsurface ocean is observed
because the heat entering the subsurface by horizontal
advection is no longer released to the atmosphere in the 
area where convection is suppressed. The nonconvective part of the box is therefore 
responsible for the rise in the box mean temperature. After 5 years convection is completely 
suppressed by the flow of relatively fresh water entering the surface box. 
Shortly thereafter, the anomalous flow of fresh water breaks down 
(see also Fig. \ref{h3.wea.over}), and the advection of salt becomes for a short time 
even positive.  The resulting rise in surface salinity quickly re-enables convection. 
As will be discussed below, the positive feedback
of convection in an ocean with relatively warm and saline water beneath a fresh and cool
surface layer now results in a rapid growth of the convective area.
This rapid growth is revealed by the sharp rise in the convective salt source for 
the surface box from 6 to 8 years. The convective area now starts to decrease again 
due to the cooling of the subsurface box and the increasing advection of 
fresh water into the surface box.

\section{The Mechanism \label{h3.wea.sec.mech}}

In this section we will explain the mechanism of the oscillation. 
The processes responsible for the expansion and retreat of the 
convective area are studied.
By definition, fluctuations in convective activity 
are caused by changes in the vertical temperature and salinity
gradient.  We focus on the surface salinity and the subsurface 
temperature because these contribute most to the variations 
in the density gradient.
In the first subsection we will consider the role of the 
subsurface temperature and surface salinity. We will 
argue that, with a steady ocean circulation that does not respond
to the fluctuations in convective activity, convection 
is self-sustaining and that changes in the subsurface temperature
and surface salinity do not cause oscillatory behavior. 
We therefore propose that ocean dynamics play a role. 
In the second subsection we will show in a conceptual model 
that changes in the surface advection due to the ocean 
dynamics can give rise to oscillations in
convective activity. Finally in the third and the 
fourth subsection we will investigate
the dynamic reponse, and the associated changes in 
advection, during the convective and the nonconvective
phase, respectively. 
  
\subsection{The expansion and retreat of the convective area.}

The role of the subsurface temperature can be summarized as follows. 
First consider the retreat of the convective area. 
Convection is suppressed only after the heat anomaly contained 
in the subsurface ocean is released to the atmosphere. 
With a warm subsurface ocean convection continues because 
of the unstable temperature gradient. In this case 
a typical heat anomaly is released in two year time, where we used a layer thickness
of 400 m, a temperature anomaly of 2 $^o$C, and an 
atmospheric heat flux of 50 W m$^{-2}$. Similarly, the warming of the subsurface 
during the nonconvective phase is a precondition for the triggering of convection. 
This warming of the subsurface during the nonconvective phase
is somewhat slower than the cooling during the convective phase,
but still relatively fast. 

When convection starts in the northeastern corner, convection mixes
up a large amount of heat and salt to the surface. 
The heat is rapidly lost to the
atmosphere, whereas the surface salinity remains high, and becomes
a source of salt for the surrounding ocean. The mixing of salt to the surface
and the subsequent horizontal spreading due to advection and diffusion,
cause -- in concert with the relatively low influx of fresh water
by advection -- a rapid expansion of the convective area. 
Due to the convective feedback, the expansion of the convective area
even continues a short time while the advection of fresh water is already 
increasing.
Thereafter, the net advection of fresh water into the convective 
area increases due to the increasing southeastward surface velocity
and the increased surface salinity gradient. The increased advection
of fresh water causes a reduction of the convective activity, and eventually
establishes a complete suppression of convection. 

In the appendix we show that convection is locally self-sustaining.
With a steady, not responding to the convective activity, circulation
convective activity will not oscillate. The mechanism proposed 
in Yin (1995, see also section \ref{h3.sec.comYS95}) 
does not operate in that case: Within a reasonable parameter regime,    
the subsurface warming when convection is off 
is not large enough to trigger the onset of convection; when convection is
on, the surface freshening due to the increased surface salinity
gradient is not large enough to suppres convection. 
Instead locally two states -- a convective and a nonconvective -- 
exist and hysteresis occurs (see also LH94, LH96). 
This leaves us with the question why the system oscillates. 
We propose that the ocean dynamics, and especially the
changes in the surface advection of salt caused by the ocean dynamics, 
play an important role. In the next paragraph we will first investigate the
role of ocean dynamics in a conceptual model. Thereafter, we will investigate 
it in the ocean model.

\subsection{A conceptual model for the oscillation}

The conceptual model is shown in Fig. \ref{h3.simple}.
It consists of two boxes: a surface box with salinity
$S$ and a subsurface box with temperature $T$. 
The temperature of the surface box and the salinity
of the subsurface box are kept constant in time. 
Convective mixing with a strength $C$ occurs between
the boxes. Furthermore, a polar overturning cell with 
strength $O$, which advects relatively fresh water into
the convective area, is shown. The system is linearized
around a mean convective state, and 
$T$, $S$, $C$ and $O$ are therefore anomalous variables. 

Let us first consider the effect of the subsurface 
temperature in isolation. 
The subsurface temperature will rise
when the convective activity is low  due to the advection
of warm water by the meridional overturning,
and fall when the convective activity is high:
\[ \frac{dT}{dt} = - \theta C, \]
where $\theta$ is a positive constant.   
The convective activity on the other hand is dependent
on the subsurface temperature. It will be high when
the subsurface temperature is high:
\[ C = \kappa T, \]
where $\kappa$ is a positive constant.
Solving this system yields a damped exponential.
The mechanism of the decay is simple:
convection causes a cooling of the subsurface,
which reduces the convective activity further. 
This feedback will be called the convective temperature
feedback. 

We continue with the surface salinity.
If we consider the effect of convection in isolation
a growing exponential is obtained in a similar way. 
In words, convection mixes salt up to the surface;
the resulting increase in surface salinity
destabilizes the stratification further and
causes an increase in convective activity.
This feedback will be called the convective 
salinity feedback.

Within this conceptual model, the convective temperature
and salinity feedback generate 
decay and amplification of the convective activity, respectively.
They are unable to generate oscillatory behavior.
This is because there is no time lag between the 
convective fluxes and the convective activity.
The essential time lag is provided by the advection 
of relatively fresh water into the convective area 
in response to changes in convective activity. This 
advective feedback can be described as follows.
The strength of the polar cell increases
when the convective activity is anomalously high:
\[ \frac{dO}{dt} = \zeta C. \]
Furthermore, the advection of relatively fresh water into the convective
area by $O$ is represented by
\[ \frac{dS}{dt} = -\gamma O, \]
and the convective activity is related to the surface
salinity by
\[ C = \sigma S. \]
In these equations $\zeta$, $\gamma$ and $\sigma$ are all
positive constants. 
Solving this system yields a second order 
ordinary differential equation:
\[ \frac{d^2O}{dt^2} + \zeta \gamma \sigma O = 0. \]
The solution of this equation is a harmonic oscillation
with a time scale 2$\pi$/$\sqrt{\zeta \gamma \sigma}$.
From the data of the ocean model we estimate $\sigma = 1$ (psu)$^{-1}$ 
-- the convective activity is normalized --, 
$\zeta$ = 3 Sv yr$^{-1}$, and $\gamma$ = 0.1 psu 
(Sv yr)$^{-1}$. This gives a time scale of 11 years.
The estimates for the parameters are obviously not very precise, but the
(inter)decadal time scale seems nevertheless robust. 
The physical picture of the oscillation is:
\begin{quotation}
 \ldots weak convective activity $C$ $\rightarrow$ slow overturning $O$
 and weak advection $\rightarrow$ high surface salinity
 and strong convective activity $C$ $\rightarrow$ fast overturning $O$ \ldots
\end{quotation}
The systems oscillates because there is a phase lag between the 
convective activity and advection of fresh water into the convective
area. 
This phase lag is clearly shown with the retreat of the
convective area. The maximum in convective activity
occurs between 10 and 12 years -- in Figs. \ref{h3.wea.budget} c,d
the convective fluxes are maximal at t = 10 years; the convective
fraction $q$ in Fig. \ref{h3.wea.depth} is maximal after 12 years.
The  maximum in the advection of fresh water in Fig. \ref{h3.wea.budget} c
occurs approximately one quarter period later, at t = 15 years. 
The phase lag is however not so evident when 
the convective activity reaches its minimum at t = 5 years: the maximum in the advection 
of salt at t = 6 years in Fig. \ref{h3.wea.budget} c 
only lags the minimum of convective activity about one year. 
This is due to the occurrence of a fast boundary wave,
which will be discussed in the following.

In the next two paragraphs we will focus on the dynamical response of the ocean model,
in particular the dynamical response of advection, during two phases of the oscillation. 
In the first paragraph we will focus on the convective phase, and in particular the 
end of the convective phase during which the convective area retreats. 
In the second paragraph we will focus on the nonconvective phase. 

\subsection{Advection during the convective phase}

To complete the picture of the mechanism of the oscillation,
the relation between convective activity and the net advection of fresh water into 
the convective area has to be established. In this section we will
discuss this relation during the convection phase. We will focus on the 
period during which the convective area retreats. 

The net advection of fresh water is determined by both the surface salinity gradient and 
the strength of the flow. The surface salinity gradient is established by two effects. 
First, precipitation creates a pool of fresh water outside the convective area mainly 
in the northwestern corner of the basin. Second, convection is responsible for a positive 
salinity anomaly in the convection area.
In the convective phase both the surface salinity gradient 
and the southeastern surface flow increase. 
In order to estimate the contribution of changes in the strength of the
surface flow and changes in the surface salinity gradient seperately, 
we divided the surface velocity into 
a time mean $v_m$ and an anomalous part $v'$. 
We computed the net advection by the mean velocity 
and the anomalous velocity. The changes in the advection of fresh
water by the mean flow $v_m\cdot\nabla S$ represents the changes 
due to surface salinity changes only.
We denote the response of $v_m\cdot\nabla S$ to anomalous convective activity shortly
by the {\em static} response of advection, because it is the response in an ocean with fixed, 
not responding to anomalous convective activity, velocities. 
The net advection by the anomalous velocity $v'\cdot\nabla S$
is the {\em dynamic} response of advection, which is due to changes in 
the strength of the flow.
 
We concentrated on two times, t = 0 and t = 4, during the retreat of the convective area.
At t = 0 the convective area is still relatively large. In Fig. \ref{h3.wea.fresh0} a,b the SSS and 
the convective activity are shown. In Fig. \ref{h3.wea.fresh0} c,d we plotted $v_m$ and $v'$, 
and the horizontal advection due to these velocities: $v_m\cdot\nabla S$  and $v'\cdot\nabla S$. 
Both $v_m$ and $v'$ advect fresh water into the southern part of the convective area; 
the contribution of $v_m$ to the advection is about about two to four times larger than the contribution 
of $v'$. The southeastern direction of the mean flow $v_m$ is
due to a combination of the windstress, which mainly forces the southward component of the flow, and the 
density gradient caused by dense water trapped along the northern boundary of the basin, 
which mainly forces the eastward component of the flow.  
For t = 4 the convective area is only small with deep convection near the northeastern corner.
As shown in Fig. \ref{h3.wea.fresh4} the anomalous southward flow clearly points into 
the convective area now, whereas at t = 0 years the anomalous flow was mainly parallel with 
the convective border. 
The horizontal flow is consequently more efficient in causing a freshening 
of the convective area. Both $v_m$ and $v'$ now contribute almost 
equally to the advection into the 
convective area. 

It is shown that the dynamic response of the advection of fresh water
is large at the end of convective phase. 
During most of the time however the static response dominates. 
In addition, horizontal diffusion (see YS95) has approximately 
the same effect as the static response of advection. 
Because the dynamic response of advection seems to be so small 
compared to the static response and horizontal diffusion, 
it could be argued that the dynamic response is not essential. 
The results of our conceptual model indicate that this is not the case. 
The formulation of the advective feedback in our conceptual model 
can be considered to represent the dynamic response of advection linearized 
around a mean salinity gradient. The phase lag between the advection 
of fresh water and the convective activity causes the oscillatory behavior. 
Since convection is a fast process, the horizontal salinity gradient
is almost in phase with the convective activity. The static response
of advecton is therefore approximately in phase with the convective
activity, and does not give rise to oscillatory behavior. In the following
we will therefore concentrate on the cause of the anomalous velocities. 

In order to get some feeling for the  response of the circulation to 
changes in convective activity we first consider the following 
idealized situation. 
Suppose that we have the following idealized situation with uniform anomalous 
convective activity in a certain area of the ocean and that this 
convective activity causes an uniform 
density anomaly in this area.
Further, assume that advective processes did not yet displaced the 
density anomaly. This situation is schematically depicted in Fig.\ref{h3.wea.sketch} a.
If we assume geostrophy, which is nearly perfectly fulfilled on the scales considered, 
the anomalous surface velocity associated with the density anomaly is a cyclonic circulation 
along the boundary of the convective area. At depth the direction of flow is reversed. 
Upwelling at the western boundary just north of the convective area, 
and downwelling just south of the convective area now close the circulation.
In this situation, the anomalous surface flow would therefore be parallel with the convective border 
and would not advect fresh water into the convective area. The dynamic response of advection 
is therefore zero in this situation. 

It is obvious that at the end of the convective phase 
our model is not in the idealized situation sketched above. At t = 4 years the 
the anomalous velocities are clearly directed into the convective 
area. Based on data of the ocean model, we sketched in Fig. \ref{h3.wea.sketch}b
the situation that is more appropriate for the end of the convective phase. 
In this situation horizontal advection caused a displacement 
of the density anomaly at depth. This is for example visible at a depth below 1000 m (not shown) 
where a dense, negative temperature anomaly is trapped in the northeastern corner of the basin. 
Associated with the density anomaly is also a surface circulation that is directed into the 
convective area (see also Fig. \ref{h3.wea.fresh4}e below).
Besides horizontal advection, also vertical advection plays a large role
in causing anomalous densities. Anomalous downwelling in a stratified ocean
causes a negative density anomaly; similarly, anomalous upwelling causes a 
positive density anomaly.  Therefore, the anomalous downwelling  
at the eastern boundary just south of the convective area in Fig. \ref{h3.wea.sketch} a
causes a negative density anomaly.
Anomalous upwelling at the border of the convective area similarly creates a positive 
density anomaly north of the convective area. The importance of upwelling in creating horizontal density 
and thus velocity anomalies has also been appreciated by Moore and Reason (1993).
In our case, the anomalous upwelling near the polar boundary during the convective 
phase, which is for example revealed by the pictures of the meridional streamfunction, 
causes a positive density anomaly at the northern boundary. This density anomaly causes an 
easterly surface velocity into the convective area.   
 
In order to quantify the role of density changes at different depths in causing 
surface velocity anomalies, we used the following diagnostic tool. Anomalous velocity 
fields are computed from the anomalous density field using the assumption that the ocean 
circulation is in geostrophic balance.  Because the ocean 
is to a very high degree geostrophic the anomalous velocity fields 
obtained from this procedure and the observed 
anomalous velocity field are almost indistinguishable.  
Then we computed the anomalous surface velocity due 
to density anomalies at different levels of the ocean. Performing 
this procedure for the anomalous density 
below 550 m, we obtained the results as depicted in Fig. \ref{h3.wea.fresh4}e. 
It is shown that about 60 \% of the southeastward velocity near the northeastern
corner, and consequently about 60 \% of the horizontal advection 
of fresh water into the convective area is due to the density changes below 550 m. 
The oscillation is therefore not only a surface and subsurface phenomenom;
also changes at depth are important in determining the dynamical reponse
of the ocean model.  
We also performed this procedure for t = 0 years. It turned out that the major 
part of the anomalous velocities is due to the density changes in the upper 
few hundred meters of the ocean. The anomalous flow only has 
a small component into the convective area, and transported therefore only 
few fresh water into the convective area. 

\subsection{Advection during the nonconvective phase \label{h3.sec.bwave}}

\begin{figure}[h]
\centerline{\psfig{figure=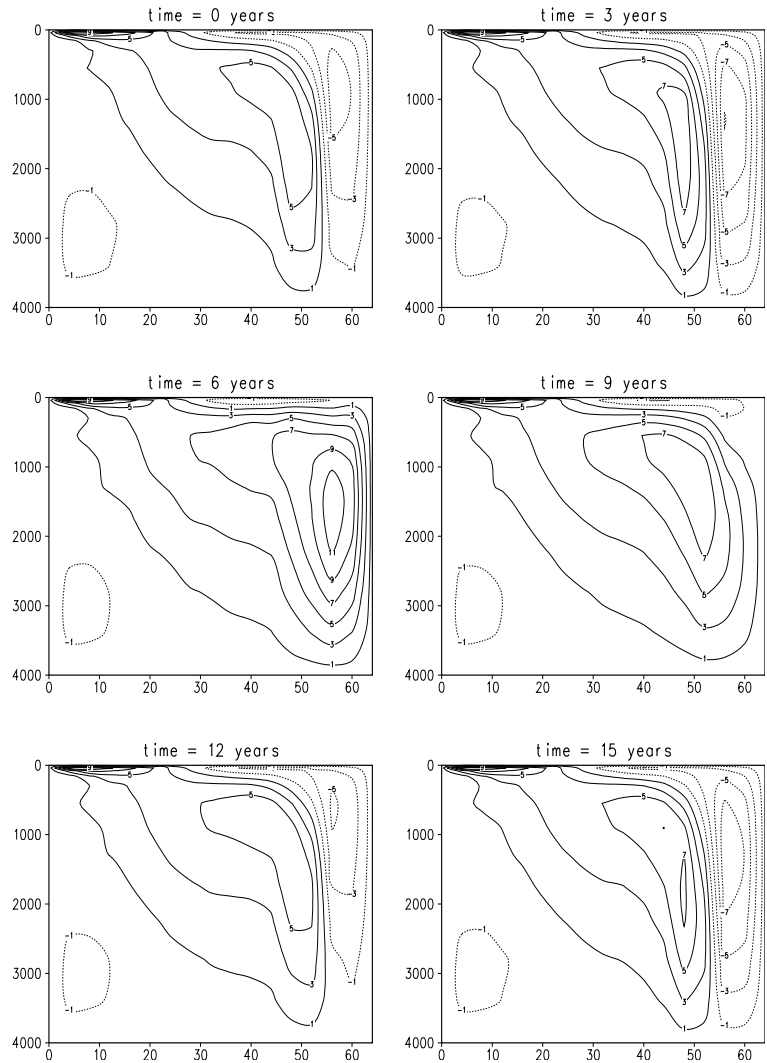,height=14cm,width=10cm,clip=}}
\center{\parbox{14cm}{\renewcommand{\baselinestretch}{1.1} 
\caption{\label{h3.wea.over}\small 
Meridional overturing at every 3 years during the oscillation.
}}}
\end{figure}

\begin{figure}[h]
\centerline{\psfig{figure=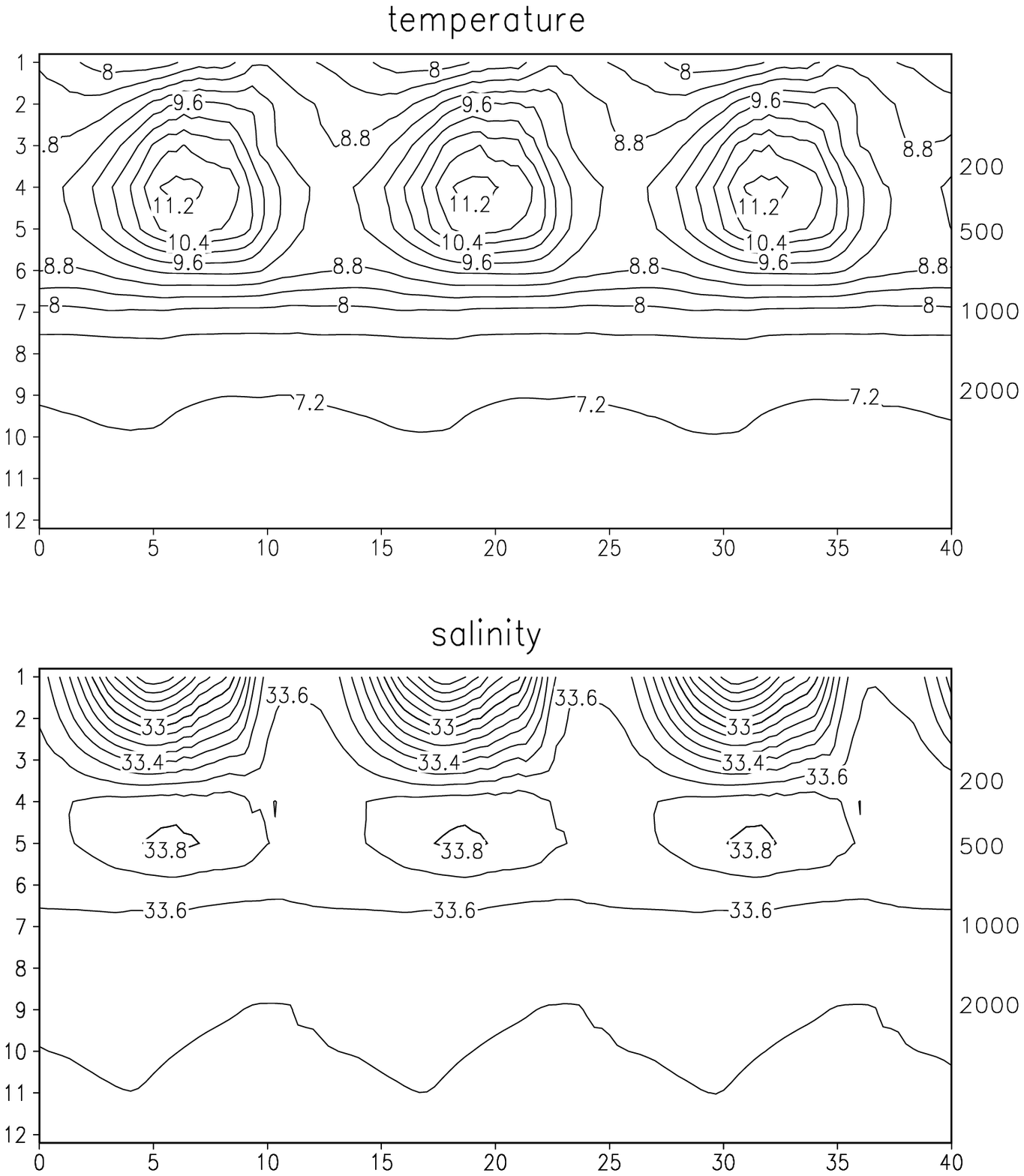,height=12cm,width=12cm,clip=}}
\vspace{0.5cm}
\centerline{\psfig{figure=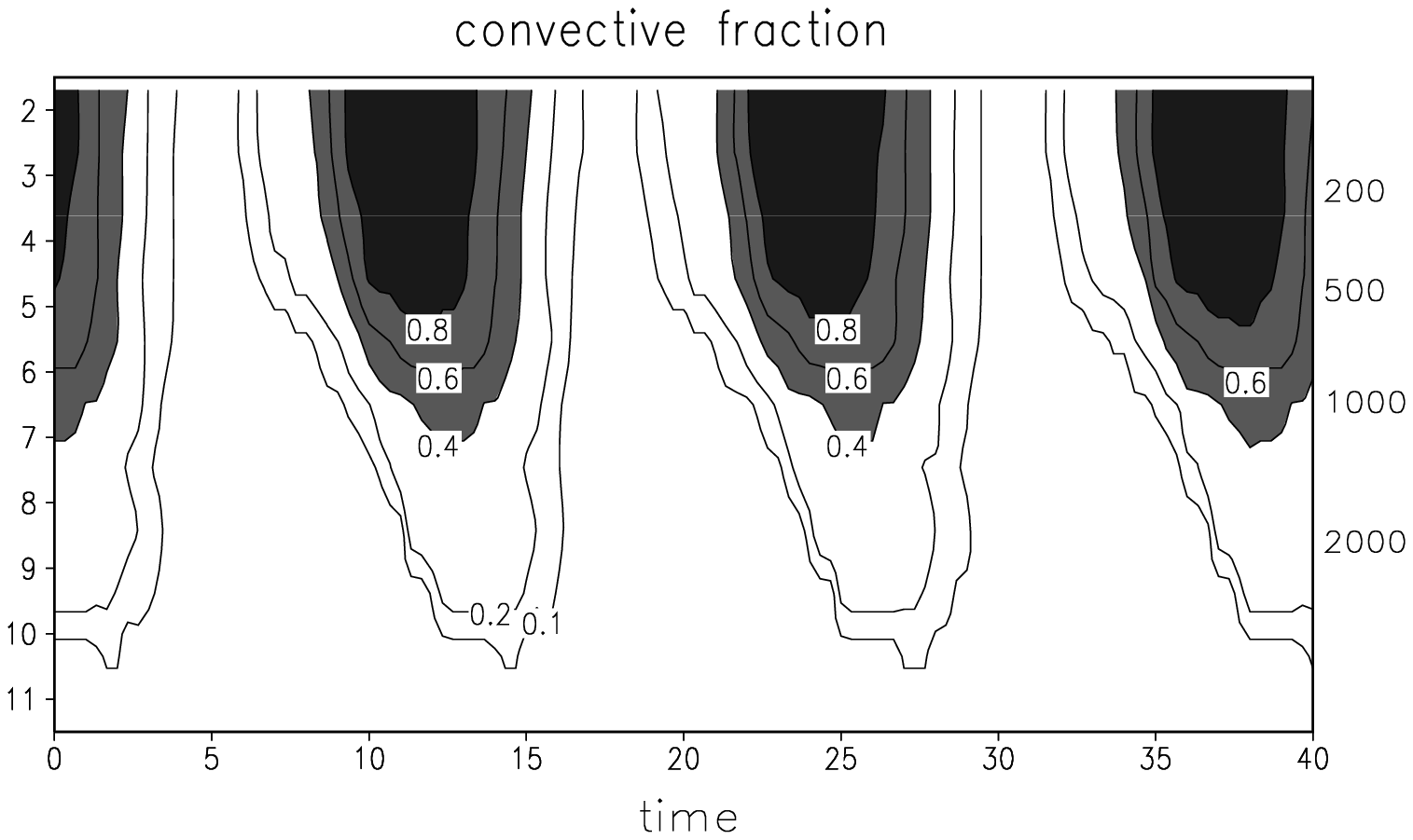,height=6cm,width=12cm,clip=}}
\center{\parbox{14cm}{\renewcommand{\baselinestretch}{1.1}
\caption{\label{h3.wea.depth}\small 
Area mean temperature and salinity and convective fraction as a function of time and depth.
At the left-hand side depth in levels, right-hand side depth in meters.
}}}
\end{figure}

\begin{figure}[h]
\centerline{\psfig{figure=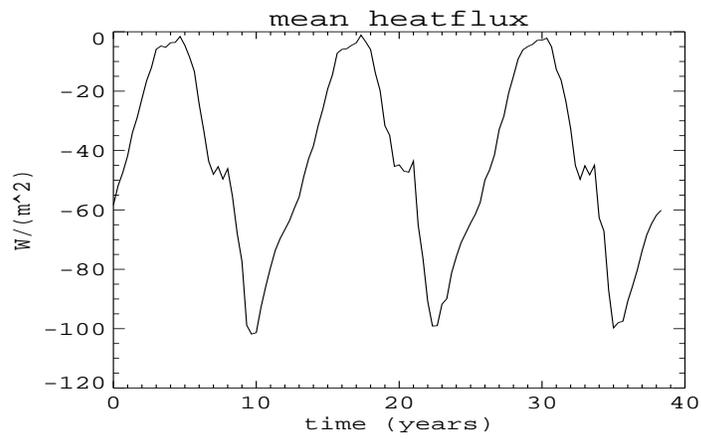,height=6cm,width=10cm,clip=}}
\center{\parbox{14cm}{\renewcommand{\baselinestretch}{1.1} 
\caption{\label{h3.wea.htflux}\small 
Mean heatflux from the atmosphere to the surface box as a function of time.
}}}
\end{figure}

\begin{figure}[h]
\centerline{\psfig{figure=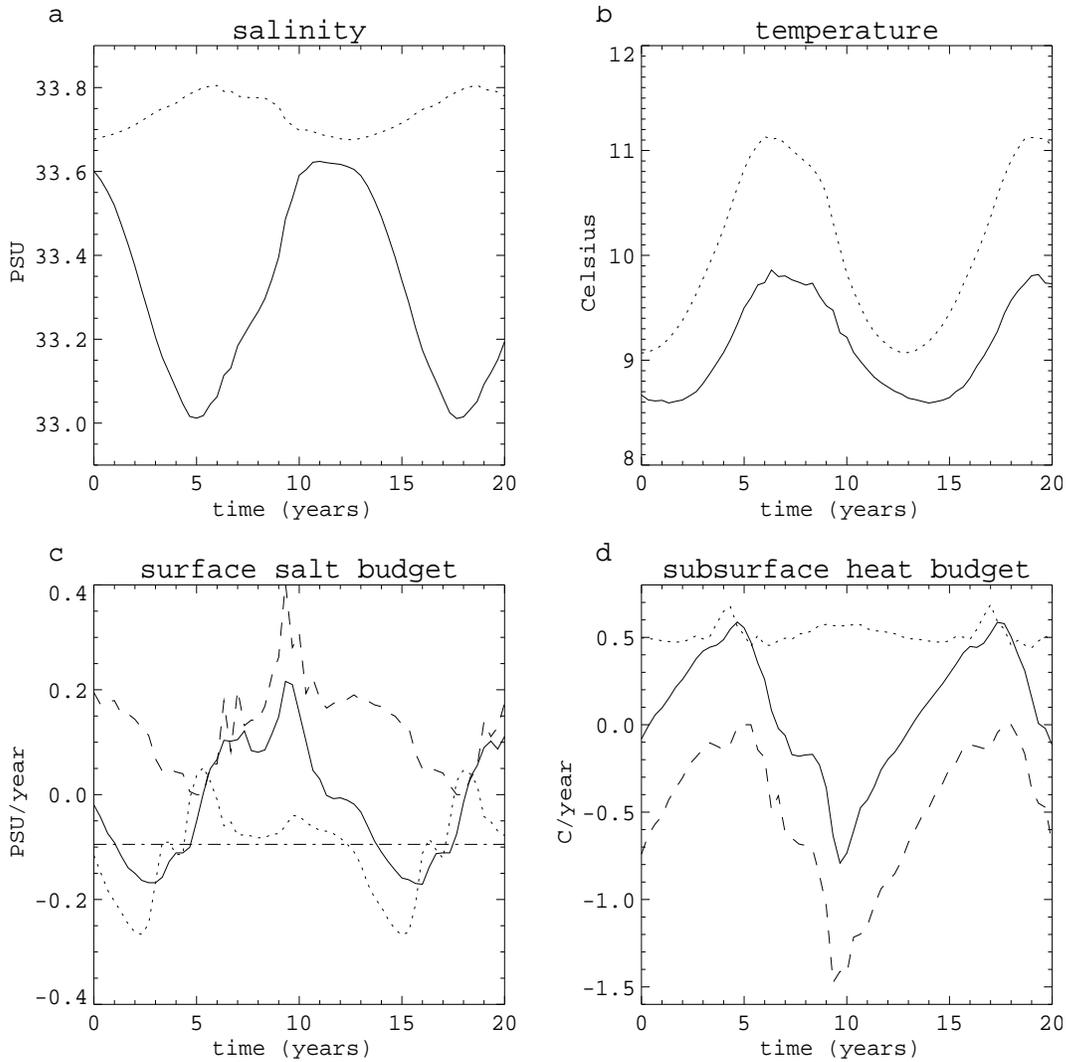,height=14cm,width=14cm,clip=}}
\center{\parbox{14cm}{\renewcommand{\baselinestretch}{1.1}
\caption{\label{h3.wea.budget}\small 
Salinity (a) and temperature (b) of the surface box (solid line) and
the subsurface box (dotted line). Salinity budget for surface box (c)
and heatbudget lower box (d) (short dash: advection, long dash: convection, 
dot-dash: net E-P, solid:
time derivative salinity (c) or temperature (d)).
}}}
\end{figure}

\begin{figure}[h]
\centerline{\psfig{figure=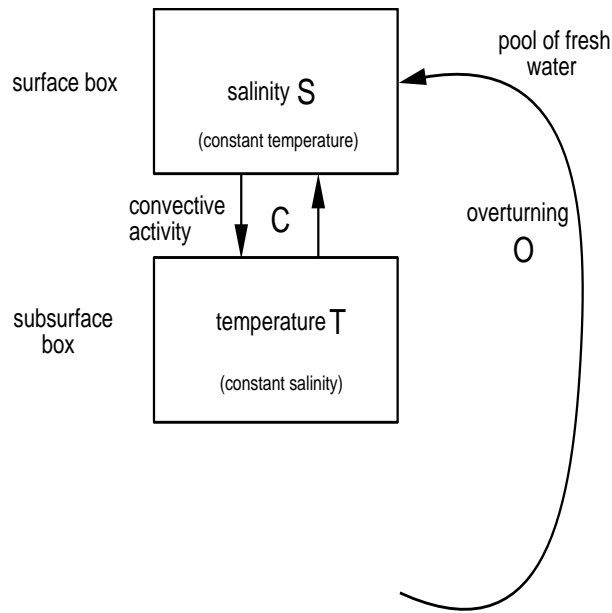,height=8cm,width=8cm,clip=}}
\center{\parbox{14cm}{\renewcommand{\baselinestretch}{1.1}
\caption{\label{h3.simple}\small
A conceptual model for the oscillation. Shown are a surface box with
variable salinity S, a subsurface box with variable temerature T, and a polar
overturning cell with strength O. The convective activity between the boxes
is denoted with C.
}}}
\end{figure}

\begin{figure}[h]
\centerline{\psfig{figure=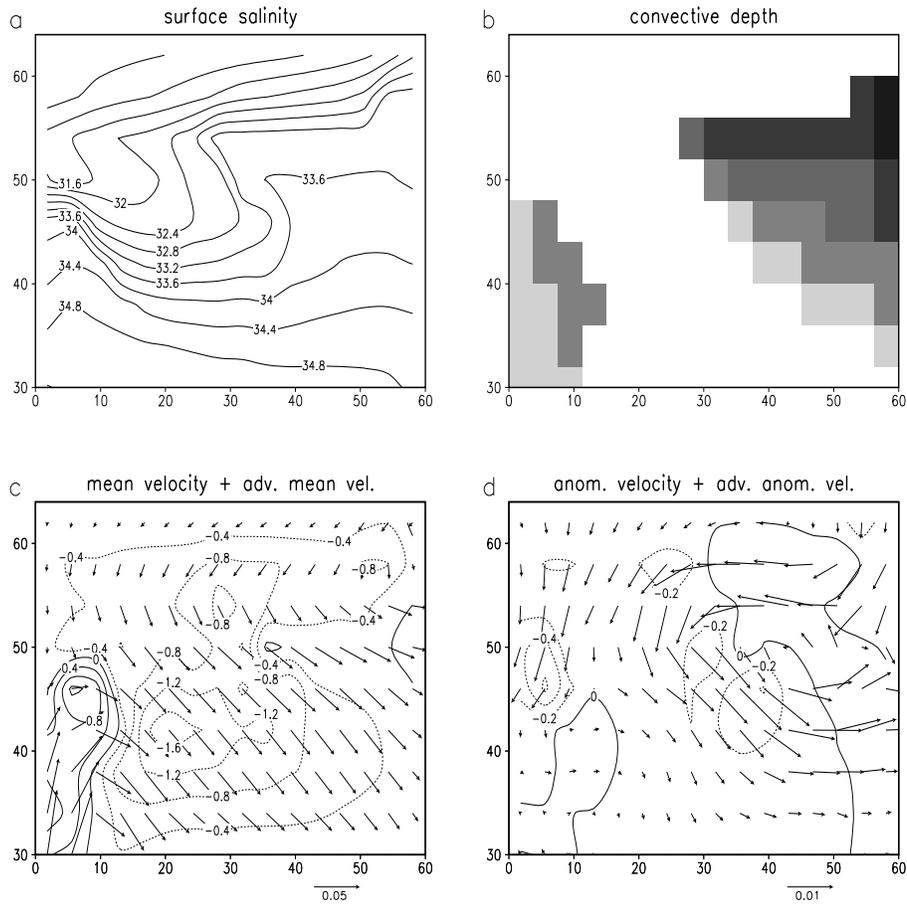,height=12cm,width=12cm,clip=}}
\center{\parbox{14cm}{\renewcommand{\baselinestretch}{1.1}
\caption{\label{h3.wea.fresh0}\small 
Surface salinity (a), convective depth (b), mean velocity ($v_m$) and advection by mean 
velocity ($v_m\cdot\nabla S$) in psu yr$^{-1}$(c), and anomalous velocity ($v'$) and advection by anomalous 
velocity ($v'\cdot\nabla S$) at t = 0 years.
}}}
\end{figure}

\begin{figure}[h]
\centerline{\psfig{figure=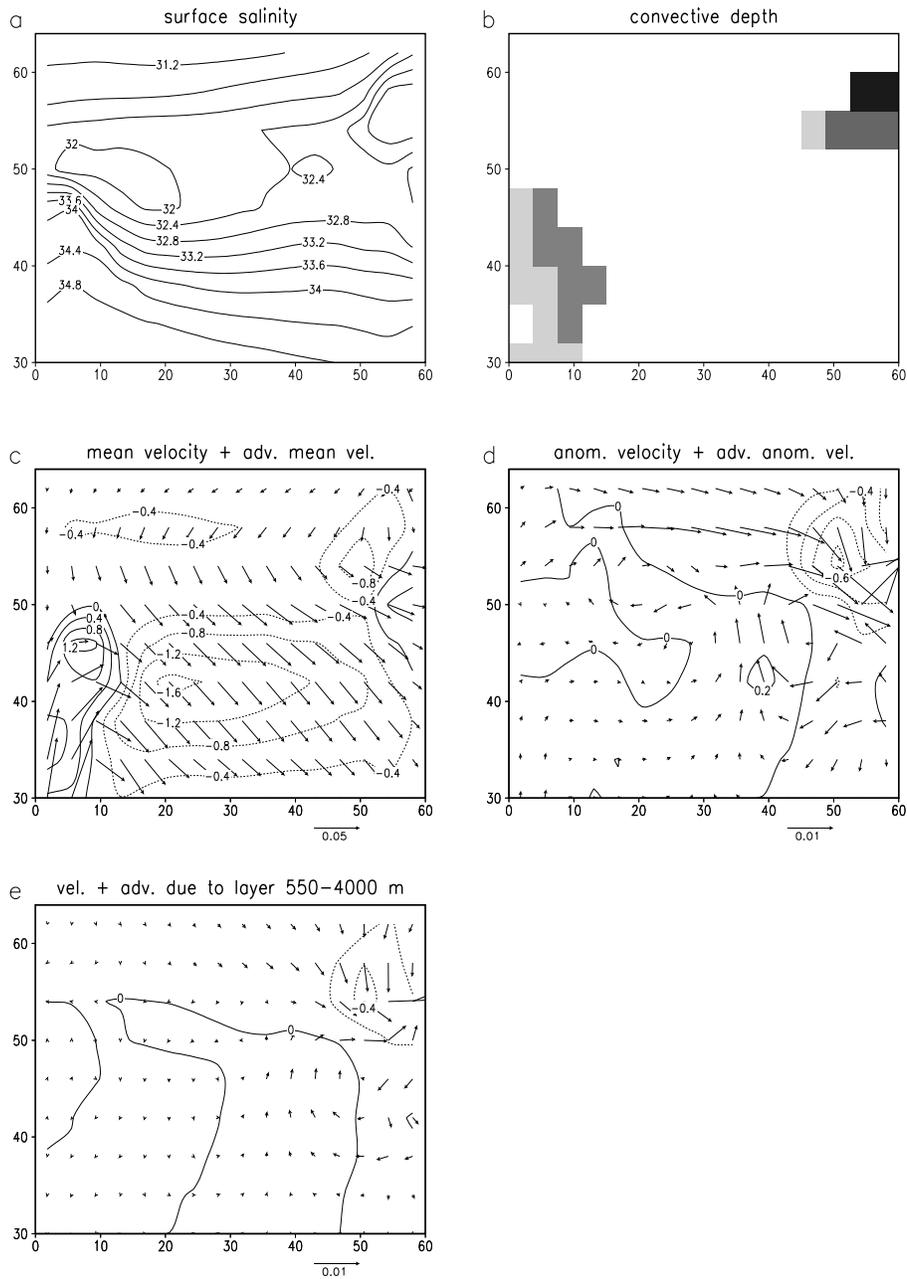,height=17cm,width=12cm,clip=}}
\center{\parbox{14cm}{\renewcommand{\baselinestretch}{1.1}
\caption{\label{h3.wea.fresh4}\small 
Similar to Fig. \ref{h3.wea.fresh0} but than for t = 4 years. 
Also shown are the anomalous velocity due to the density anomaly beneath 550 m,
and the advection due to this velocity (e).
}}}
\end{figure}

\begin{figure}[h]
\centerline{\psfig{figure=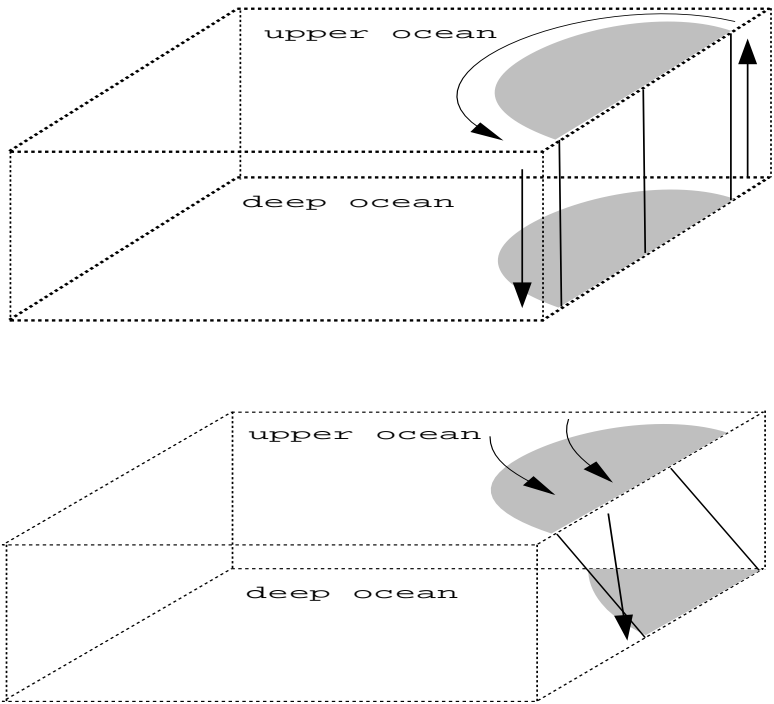,height=14cm,width=14cm,clip=}}
\center{\parbox{14cm}{\renewcommand{\baselinestretch}{1.1}
\caption{\label{h3.wea.sketch}\small 
Sketches of the relationship between convection, density anomalies, and anomalous velocity
fields. The density anomaly is depicted by the dashed area. Anomalous velocities are
shown by the arrows.
}}}
\end{figure}

It is already shown in Fig. \ref{h3.wea.over} that during 
the nonconvective phase the negative overturning cell
at the polar boundary collapses.  
This is due to the occurrence of a boundary Kelvin wave.
Kelvin waves have been observed in several studies 
of variability in coarse resolution ocean model, and some authors
claim their essential role for variability.   
(Winton 1996a; Greatbatch and Peterson 1995). Strictly speaking this wave is not 
resolved in coarse resolution ocean models, however, these models do generate a
numerical wave with equivalent characteristics (Winton 1996a). 
In a strongly stratified 
equatorial ocean the propagation speed of this wave is fast. In the weakly stratified 
polar ocean however the wave considerably slows down and could be responsible for decadal variability. 

The physics of this wave can be understood as follows. Consider an area with downwelling
at the eastern boundary in a stratified ocean. This introduces a local 
density minimum by the downward advection of lighter water. The negative density anomaly 
introduces a cyclonical circulation at depth and an anti-cyclonical surface circulation. 
The corresponding vertical circulation at the boundary with upwelling south and downwelling north of 
the anomaly, now leads to a northward displacement of the downwelling area. 
In a stratified ocean the wave therefore travels cyclonically around the basin.

\begin{figure}[h]
\centerline{\psfig{figure=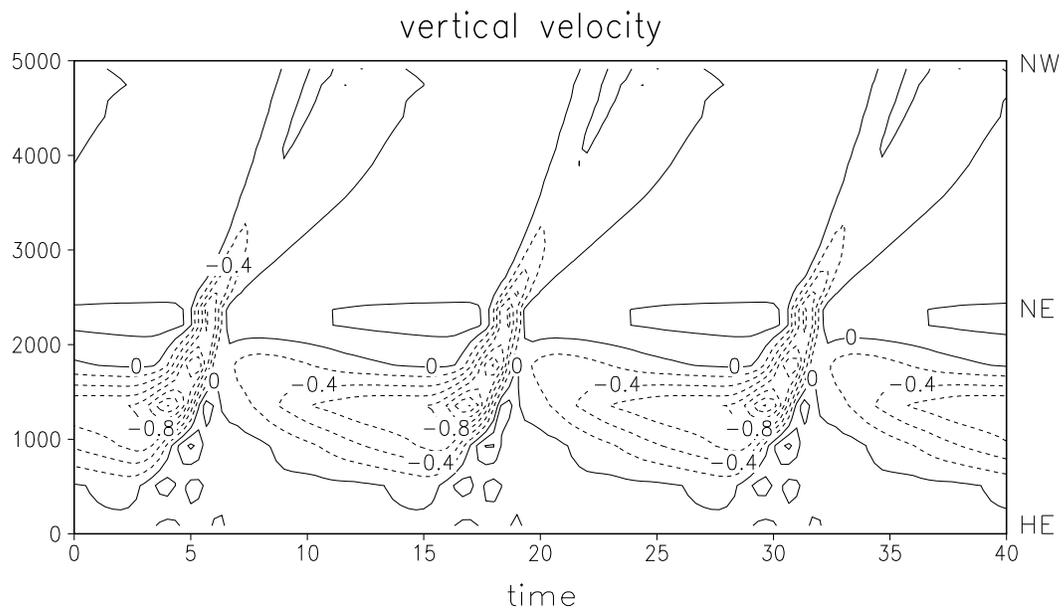,height=8cm,width=14cm,clip=}}
\center{\parbox{14cm}{\renewcommand{\baselinestretch}{1.1}
\caption{\label{h3.wea.kelvin}\small 
Time development of vertical velocity at 1000 m as a function of distance (km) along the 
boundary. HE corresponds to 32 $^o$N along the eastern boundary, NE the northeastern
corner, and NW the northwestern corner of the basin.
}}}
\end{figure}

To illustrate the propagation of the Kelvin wave, we plotted in Fig. \ref{h3.wea.kelvin}
the vertical velocity as a function of time along part of the boundary.
This part consists of the polar part of the eastern boundary 
(HE to NE in Fig. \ref{h3.wea.kelvin}) and the polar boundary (NE to NW). 
The vertical velocity occuring at the boundary is for a large part
responsible for the zonally integrated vertical mass transport as depicted in plots
of the meridional overturning.  
From t = 0 to t = 4 the downwelling 1000 km south of the polar boundary is clearly 
visible. When after 5 years deep convection at the northeastern boundary stops,
the downwelling area starts to travel cyclonically along the boundary. 
In about two years time the wave reaches the polar boundary, which is also revealed 
by the collapse of the negative, polar overturning cell in Fig. \ref{h3.wea.over}. 
Thereafter the wave travels along the polar boundary to the west, while slowly 
reducing in strength.
The time scale of the wave propagating along the boundary is about five years.

\begin{figure}[h]
\centerline{\psfig{figure=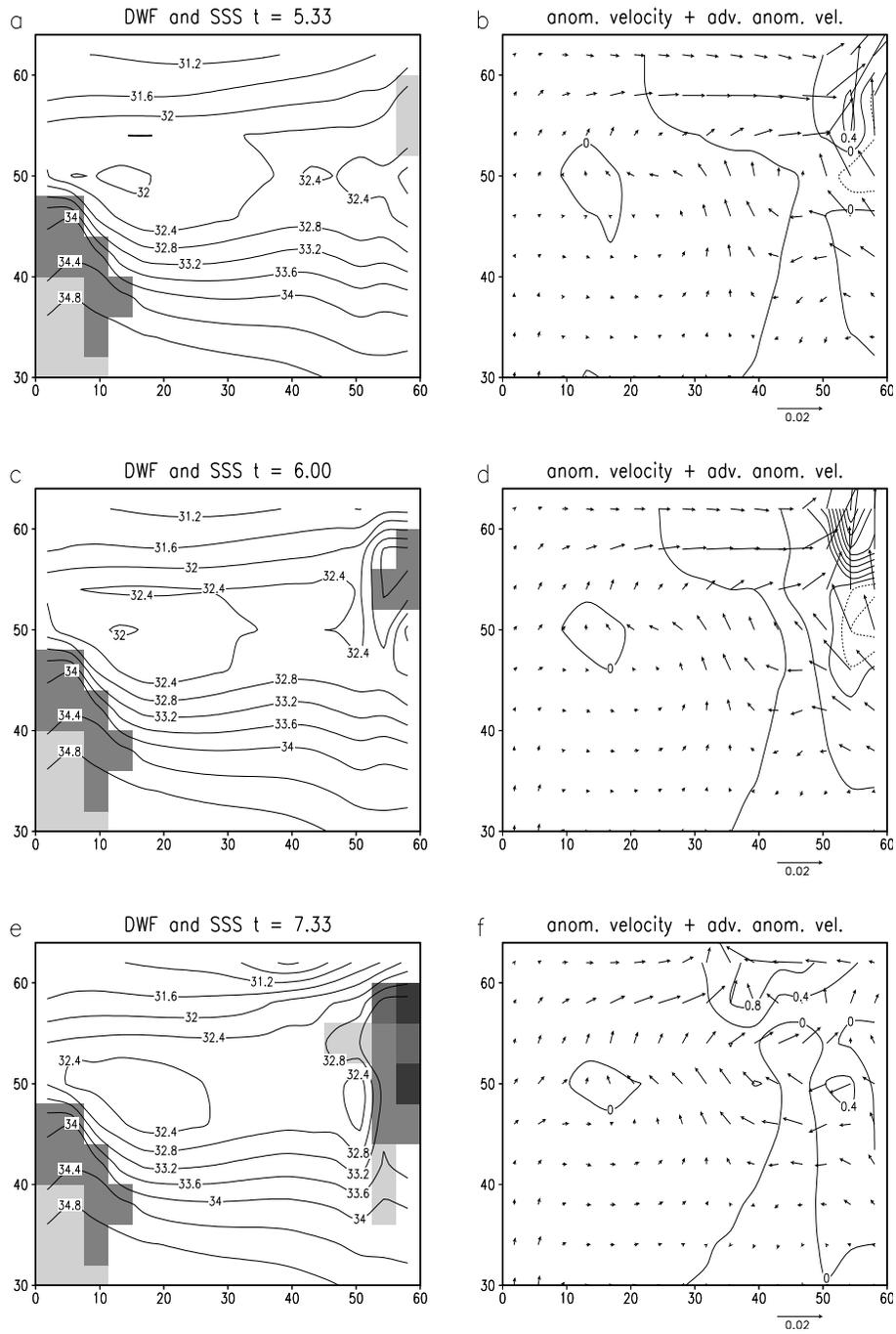,height=18cm,width=12cm,clip=}}
\center{\parbox{14cm}{\renewcommand{\baselinestretch}{1.1}
\caption{\label{h3.wea.kelvinreeks}\small 
Surface salinity, deep-water formation, anomalous surface velocity, and surface 
salt advection (psu yr$^{-1}$) due to anomalous velocity during the weak convective 
phase. 
}}}
\end{figure}

The effect of the Kelvin wave on the ocean dynamics, in particular the advection
of salt at the surface, is illustrated in Fig. \ref{h3.wea.kelvinreeks}.
Figure \ref{h3.wea.kelvinreeks} shows the surface salinity, convective depth,
anomalous velocity, and surface salt advection due to the anomalous
velocity are shown for three different times, t = 5.33, t = 6.00 and t = 7.00 years.
At t = 5.33 years, the Kelvin wave is still located at the eastern boundary, approximately
at 56 $^o$N. The anti-cyclonical surface circulation at the eastern boundary 
associated with the Kelvin wave advects relative saline water northward. This rapidly
turns convection on -- in the appendix it is argued 
that a change in the surface velocity is necessary for
the re-initialization of convection. At t = 6 years, the boundary wave is
located in the northeastern corner of the basin, and one year later it is about halfway the 
northern boundary. Associated with the propagation of the Kelvin, the strong 
eastward surface flow along the northern boundary of the basin turns into
a weakly western flow. With the relatively weak (south)eastern surface flow, 
or even westerly flow, the convective area expands rapidly. 
Thereafter, the eastern surface flow starts to increase
again. To summarize, the role of the Kelvin wave is mainly twofold: First, 
the northward surface flow associated with the wave re-initializes convection 
rapidly after its suppression. Second, it resets the ocean from a state 
with the strong, (south)eastern surface flow to a state with a weak surface 
flow, so that the convective activity is able to expand rapidly.
The dynamical response of the ocean during the nonconvective phase is
therefore dominated by the Kelvin wave. On the other hand,  
the role of Kelvin waves during the convection phase seems to be minor. 
We could not find any evidence of wave activity during this period.  
 
The suggestion of Winton (1996a) that the Kelvin wave, when it
reaches the energetic western boundary current, 
influences the overturning and therefore modifies
the advection of heat and salt from the south into the convective area seems not likely.
In the previous section it was shown that the suppression of convection is mainly
due to the anomalous advection of fresh water from the northwest into the convective 
area. Furthermore, Wintons hypothesis implies that the suppression of convection
is due to the slow down of the positive overturning circulation 
(because the overturning advects salt into the convective area) while 
actually an increase in strength of the positive overturning cell is observed
between t = 0 and t = 4 years.
We therefore conclude that boundary waves play a role in the sense that 
they cause the rapid re-initialization of convection and reset the strong (south)eastern
surface flow at the end of the convective phase, which enables the convective 
area to expand. 
There are no indications that the suppression of convection is due to boundary waves. 
The underlying physics responsible for the propagation of the wave are however important
during the whole cycle of the oscillation.
Density anomalies due to anomalous upwelling and downwelling at the boundary determine
for a large part the dynamic response of the advection of fresh water into the convective area.
Boundaries therefore play an important role in the mechanism of the oscillation.

\subsection{Experiments with fixed velocity fields}

\begin{figure}[ht!]
\centerline{\psfig{figure=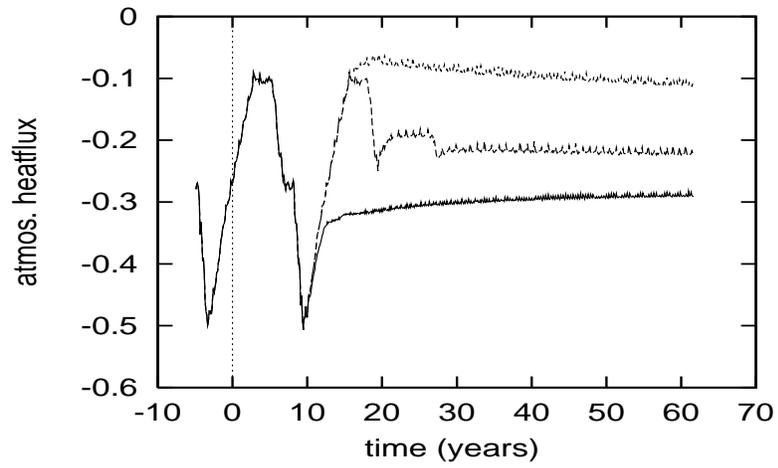,height=6cm,width=10cm,angle=270}}
\vspace{0.5cm} 
\center{\parbox{14cm}{\renewcommand{\baselinestretch}{1.1}
\caption{\label{h3.yin.heatfl}\small 
Integrated atmospheric heat flux as a function of time. Upper curve:
ocean velocity fixed after 15 years, middle curve: ocean velocity 
after 19 years set to the mean velocity during the first
two cycles, and lower curve: ocean velocity fixed after 10 years. 
}}}
\end{figure}

To confirm once more that a dynamical response of the ocean model 
to anomalous convective activity is essential, 
we performed a few idealized experiments with the ocean model. 
Thereto, we ran the ocean model with stationary velocity fields.
For the velocity fields we used the mean velocity field and also
two instantaneous, diagnosed velocity fields. As a measure of the convective activity we 
show the integrated heatflux in Fig. \ref{h3.yin.heatfl}.
For the first experiment we first integrated the model 
during two periods forward in time, and then switched over to the mean velocity field. 
For the other two experiments we integrated the model to t = 10 years
and to t = 15 years respectively, and then kept the velocity field fixed. 
These times are chosen because the southeastward velocity at 
the surface into the convective area are then respectively maximal and minimal.
The strength of this flow determines the time scale at which the surface salinity
of the convective area tends to approach the salinity of the northwestern corner,
which potentially (see the Appendix) could be important for the 
occurrence of the oscillation.  
With the velocity fields used a wide range of realistic relaxation
time scales is covered. 
None of these integrations, however, yielded self sustained oscillations
in convective activity on a (inter)decadal time scale.  After switching to a steady velocity field 
all runs only showed small variations on a very short time scale. 
No (inter)decadal oscillations were found. These experiments therefore
indicate that ocean dynamics play an essential role in (inter)decadal oscillations.

\subsection{Summary of the mechanism}

We start the cycle in the phase with weak convective activity.
A short time later convection is 
completely suppressed due to the strong advection of relatively fresh surface 
water from the northwestern polar boundary into the convective area. 
When convection stops the downwelling 
area just south of the convective area at the eastern boundary of the
basin begins to travel cyclonically around the basin as a boundary Kelvin wave.
The movement of the downwelling area is revealed by a collapse of the negative
overturning cell near the polar boundary.
The advection of relatively fresh water from the polar 
area into the convective area breaks down, and convection starts again at the 
eastern boundary. When convection is off, the subsurface becomes
warm and saline compared to the surface due to the influx
of heat and salt at depth. In the nonconvective phase this stratification
has become very unstable -- that is, potentially convective (see LH96) -- using MBCs. 
The relatively weak advection of fresh water 
(weak southeasterly flow) and the large amount of salt mixed up by the convective 
patch at the boundary cause a rapid growth of convection. 
During the subsequent strong convective phase 
the heat contained in the subsurface ocean between approximately 160 and 600 meter deep 
is rapidly released to the atmosphere. 
The cooling of the subsurface tends to stabilize the stratification. 
At the same time the advective influx of fresh water increases. This increase is due to the 
increased horizontal salinity gradient between the convective area and the pool of fresh water 
at the northwestern part of the basin, and the increased southeastward
flow into the convective area. The anomalous flow is caused by density anomalies at depth
due to horizontal advection, and anomalous upwelling and downwelling near boundaries.
The area with convection declines, followed by a complete
shut down of convection and the propagation of the boundary wave. 

\section{Sensitivity experiments \label{h3.sec.sensitive}}

\subsection{Sensitivity to salt perturbations}

For stationary equilibrium states the 
response of the ocean circulation to surface salinity perturbations is usually large
with MBCs. Often the meridional overturning completely collapses, called a polar halocline
catastrophy (Bryan 1986). In other experiments the response is somewhat more moderate (LH96). 
We therefore tested the sensitivity of this oscillation to surface salinity perturbations. 

\begin{figure}[ht!]
\centerline{\psfig{figure=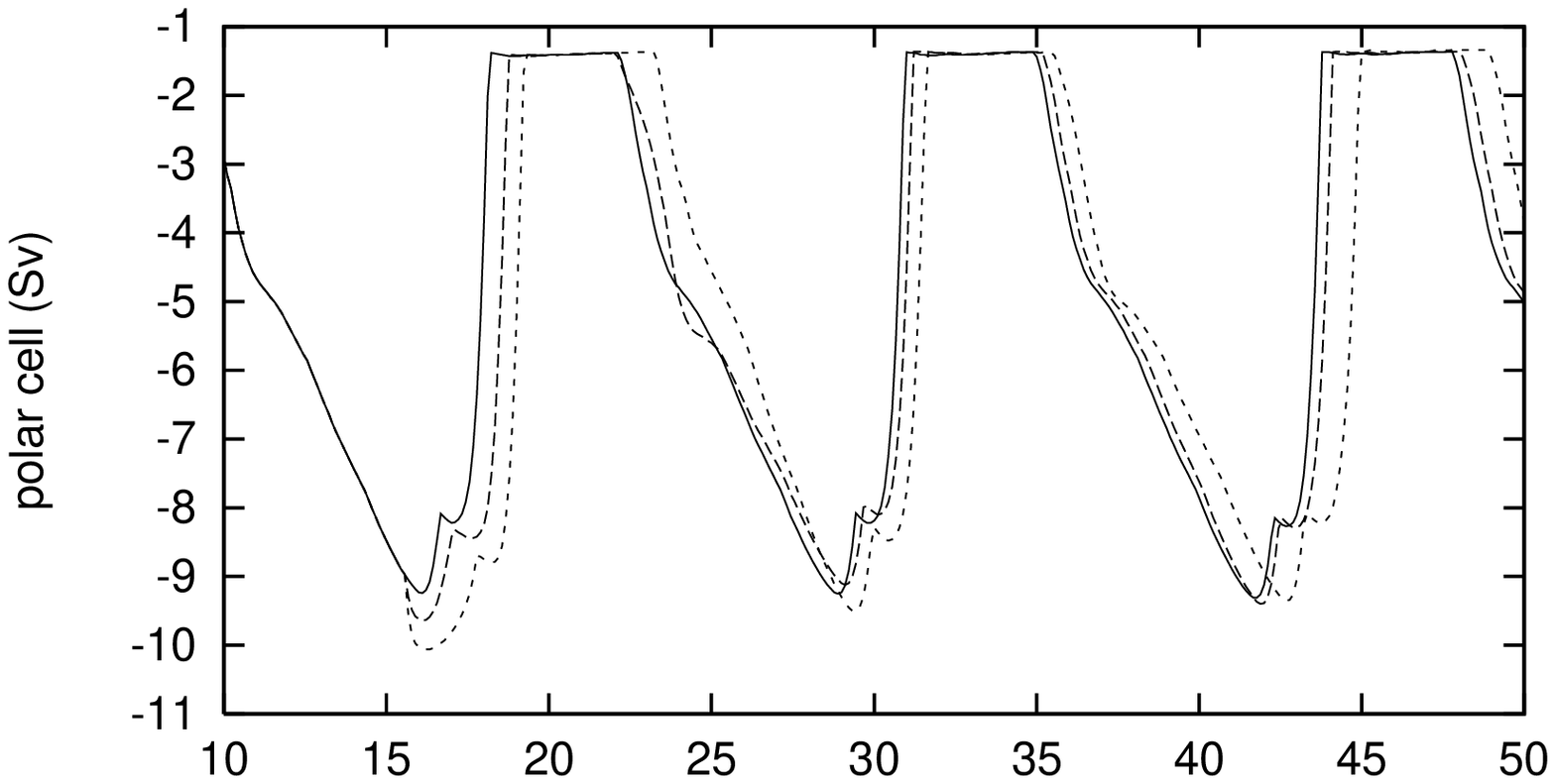,height=4.5cm,width=13cm,angle=270}}
\vspace{0.1cm} 
\centerline{\psfig{figure=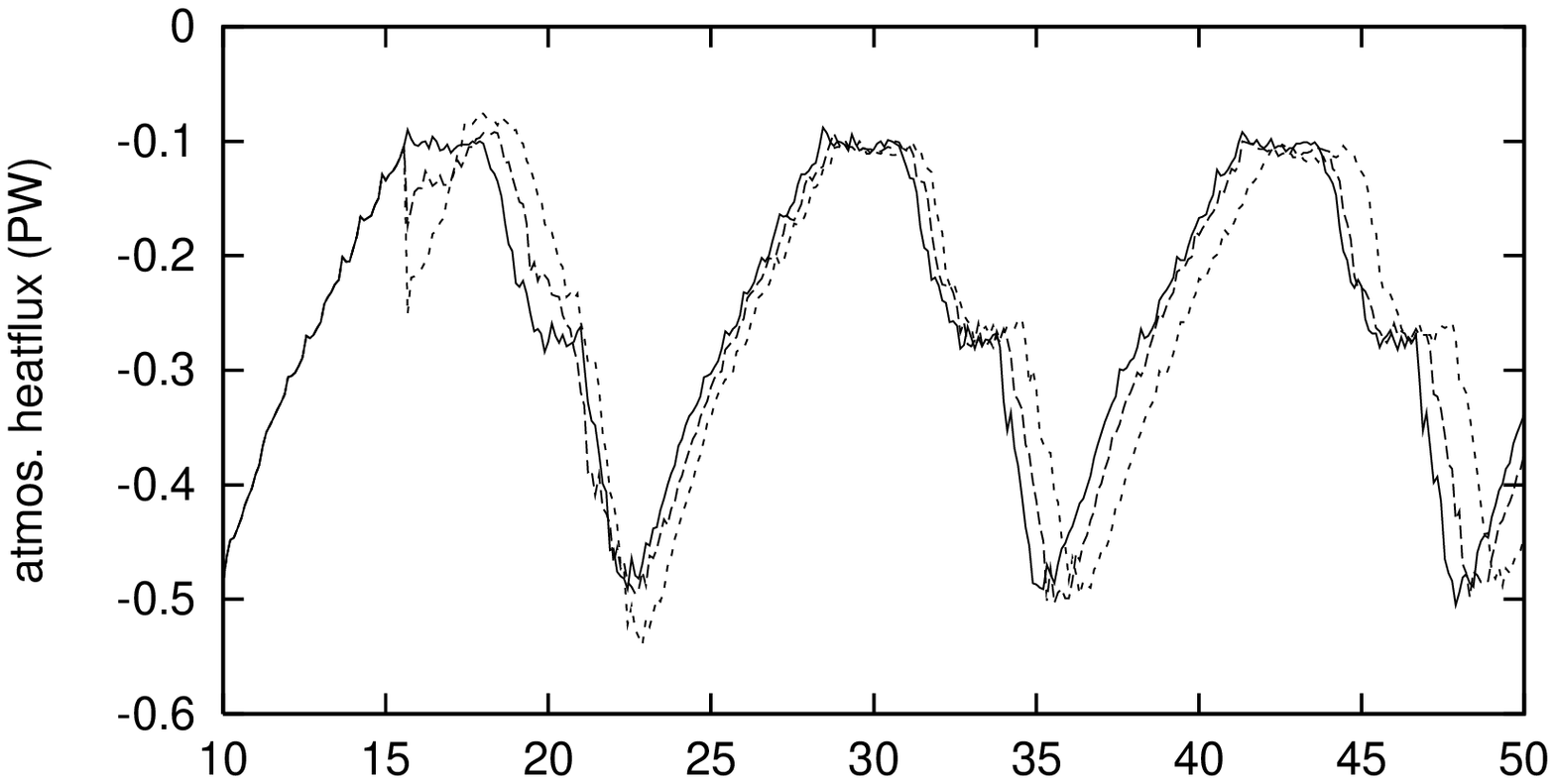,height=4.5cm,width=13cm,angle=270}}
\vspace{0.1cm} 
\centerline{\psfig{figure=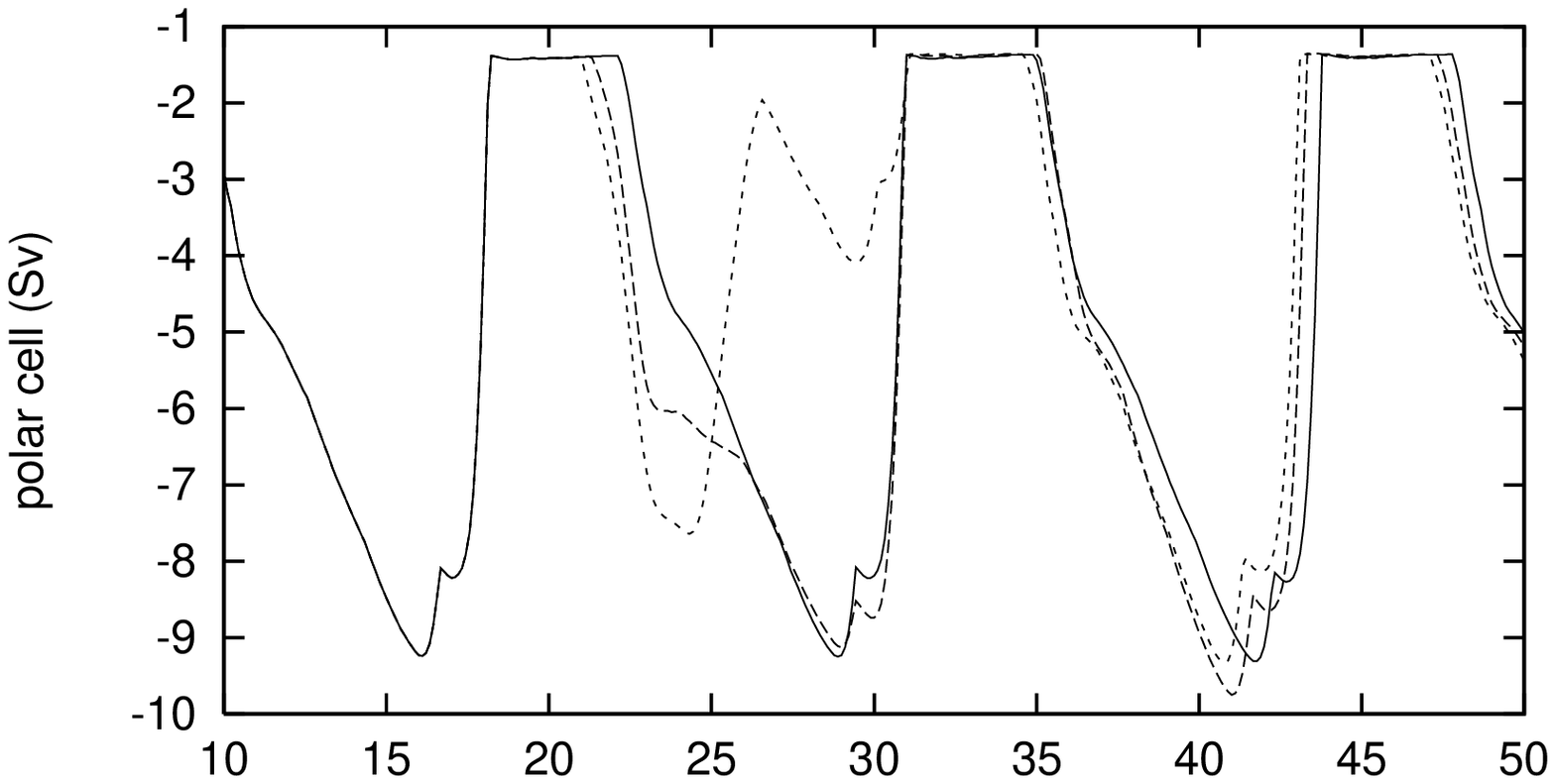,height=4.5cm,width=13cm,angle=270}}
\vspace{0.1cm} 
\centerline{\psfig{figure=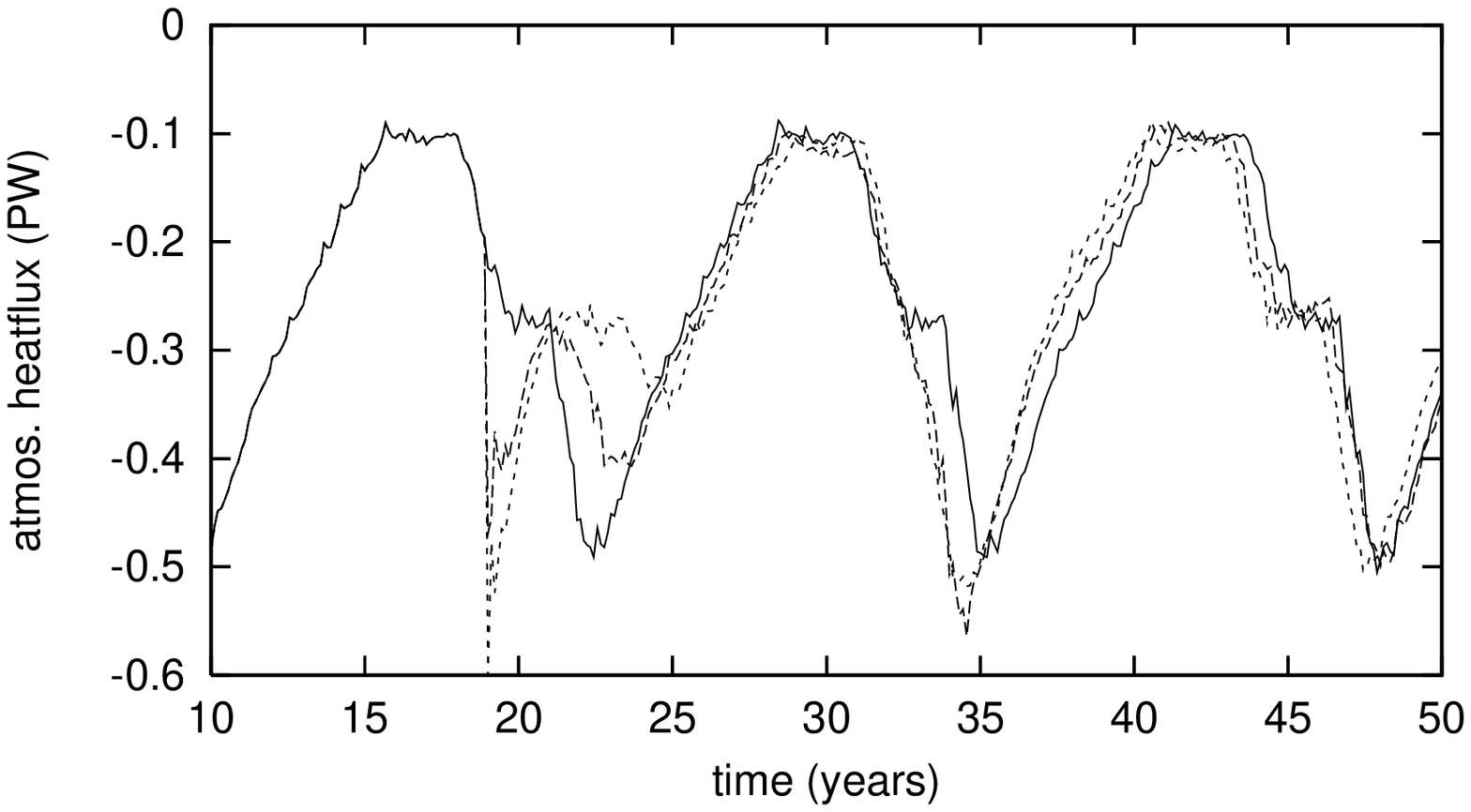,height=4.7cm,width=13cm,angle=270}}
\center{\parbox{14cm}{\renewcommand{\baselinestretch}{1.1} 
\caption{\label{h3.yin.pert}\small 
Heatflux and strength polar cell as function of time.
Upper two panels salinity perturbation after 16 years, 
lower two panels salinity perturbation after 19 years 
(solid line: nonperturbed run, long dash: 0.4 PSU, short dash: 0.8 PSU. 
}}}
\end{figure}

We perturbed the reference run after 16 years with respectively 0.4 psu and 0.8
psu the area $A_{con}$. At this moment the convective activity 
is low, the heatloss to the atmosphere is small and the polar cell is strong. 
The response of the atmospheric heatflux and the strength of the polar cell to both 
perturbations is shown in Fig. \ref{h3.yin.pert}. The perturbation causes anomalous
convective activity, which is responsible for the moderate increase in atmospheric heatflux.
In response to this, the polar cell increases in strength. An increase in the 
advection of fresh water into the convective area results. The anomalous convective 
activity rapidly disappears, and the polar cell breaks down. 
The perturbation results in a small phase shift in the order of one 
year.
We repeated this experiment and perturbed the same area with the same 
salt anomalies, but now after 19 years. 
The response of the heatflux, as shown in Fig. \ref{h3.yin.pert}, is much 
more vigorous compared to the previous experiment. 
The heatloss increases to 0.45 PW for the 0.4 psu anomaly, and to 0.60 PW for the 0.8 psu
anomaly. The more vigorous response is due to the warming of the subsurface which takes
place between year 16 and year 19. 
The negative cell now increases rapidly in strength. This causes a freshening of the convective
area, which results in a reduction of the convective activity and the atmospheric heatloss.
For the largest perturbation the decrease in convective activity even triggers 
a collapse of the negative overturning cell, for the smallest perturbation the collapse
does not take place.  
In the long term the response of the circulation to these perturbations 
is surprisingly small. Only small phase shifts in the oscillation are observed.
The reason why the perturbed runs synchronize
is not investigated in detail, but apparently it is due to 
the dynamics of the ocean. In section \ref{h3.sec.comYS95} we will argue
that with a nondynamical picture of the oscillation this behavior of the 
oscillation cannot be explained. 

\subsection{Sensitivity to the restoring constant}

The value of the restoring constant for the SST employed in Wea93 is approximately
100 W m$^{-2}$ K$^{-1}$, which is generally thought to be unrealistically high. 
In addition, the thermohaline circulation has shown to be rather sensitive to 
perturbations when a short restoring time scale is used -- a sensitivity 
which largely diminishes when longer restoring time scales are used (Power and Kleeman 1994,
Zhang et al. 1993). Therefore, we investigated the robustness of the oscillation 
with respect to the restoring timescale. In YS95 a similar experiment was performed with 
similar results.

\begin{figure}[h]
\centerline{\psfig{figure=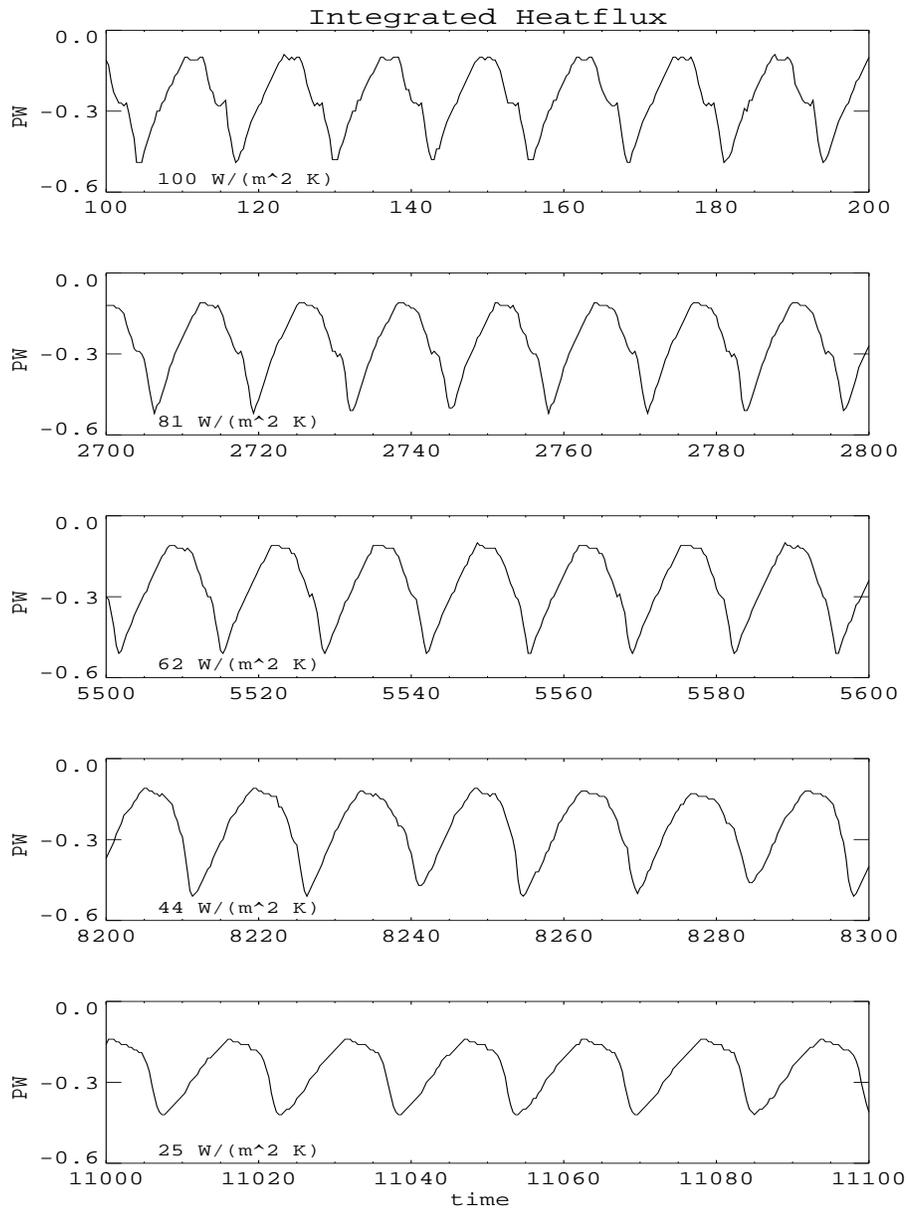,height=16cm,width=12cm,clip=}}
\center{\parbox{14cm}{\renewcommand{\baselinestretch}{1.1} 
\caption{\label{h3.wea.var}\small 
Atmospheric heat flux integrated over the polar part of the basin 
of different values of the restoring constant $\alpha$.
}}}
\end{figure}

We performed an integration with the ocean model in which the restoring 
constant is slowly reduced 
from 100 W m$^{-2}$ K$^{-1}$ to 25  W m$^{-2}$ K$^{-1}$ during a period
of 11,000 years. The slow rate of change allows the system to remain 
statistically close to equilibrium.  
We started with a state after 2000 years of the 
run in section \ref{h3.sec.weaver}. 
During the time integration the system remains oscillating.
As a measure of the oscillation we considered the integrated atmospheric heatflux
over the polar part of the basin, from 32 $^o$N to 64 $^o$N,
which is a good indication of convective activity. 
In Fig. \ref{h3.wea.var} we plotted parts of this time series 
for 5 different periods of 100 years. The value of the restoring
constant $\alpha$ during these periods is approximately 100, 81, 62,
44, and 25 W m$^{-2}$ K$^{-1}$, respectively. The results of the time periods
in between are similar.

Fig. \ref{h3.wea.var} shows that the timescale of the 
oscillation increases from 12 years to 16 years when the restoring 
constant decreases. The increase is small considering to the 
large range over which $\alpha$ varies. The increase of the time scale is due to 
several effects. First, the heat exchange between the ocean and the atmosphere
becomes less efficient, which implies that the timescale for the cooling of the subsurface
due to convection becomes longer, thus increasing the timescale of the oscillation.
Second, since the cooling becomes less efficient the impact of 
convection on the ocean density becomes less. This slows down the response 
of the ocean circulation to changes in convective activity, which results
in an increase of the time scale of the oscillation.  
In Lenderink (1996) we show in a two box model with a dynamical feedback
that the second effect is likely to be dominant.

The amplitude of the variations in the integrated heatflux remains nearly constant.  
For $\alpha$ = 25 W m$^{-2}$ K$^{-1}$ the amplitude of the oscillation 
only decreases with 40 \% compared to the reference run.
This insensitivity is surprising considering that 
$\alpha$ directly influences atmospheric heatflux. 
The insensitivity is partly related to the size of the convective area. 
An increase in the size of the  convective area compensates for the local decrease 
in heatloss, and makes the area integrated heatflux fairly insensitive to the restoring 
constant. The insensitivity can also be understood as follows.
The amplitude of the oscillation is mainly determined by the rate 
of subsurface warming, which is largely determined by the strength of the 
meridional overturning. 
Since the meridional overturning is fairly insensitive to the restoring constant, 
so is the amplitude of the oscillation.

We also performed a run with the restoring constant decreased to 
10 W m$^{-2}$ K$^{-1}$. During this run the oscillation stopped, which indicates 
that the oscillation occurs only when the subsurface ocean is able to release 
heat to the atmosphere with a relatively fast time scale.

\section{Comparison with Yin and Sarachik (1995) \label{h3.sec.comYS95}}

In Yin and Sarachik (1995) a similar mechanism
of the oscillation is proposed. They also 
argue that the increased 
advection of relatively fresh water, which is due
to the increased horizontal salinity 
gradient and a strengthening of the southeastern flow,
supresses convection. On the other hand, they argue
that the onset of convection is solely due to
the warming of the subsurface. In our opinion, however, 
the warming of the subsurface is a necessary pre-condition.
The actual triggering is due to changes in the surface
salinity. 

\begin{figure}[t]
\centerline{\psfig{figure=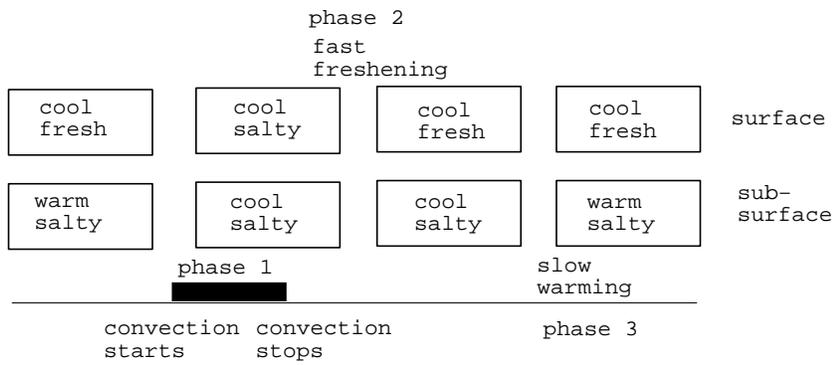,width=11cm}}
\vspace{0.5cm} 
\center{\parbox{14cm}{\renewcommand{\baselinestretch}{1.1}
\caption{\label{h3.yin.mech}\small 
Simple sketch of the mechanism of decadal oscillations by Yin 
}}}
\end{figure}

They argue that the basic mechanism of the oscillation is 
grasped in a simple box model, which is described in 
Yin (1995). 
The physical mechanism that leads to the oscillation 
is sketched in Fig. \ref{h3.yin.mech}. It is 
basically Welander's oscillation -- that is, the oscillation
results because no stable equilibrium states exist. 
In the convective state
the advection of fresh water supresses convection;
in the nonconvective state the advection of heat 
at the subsurface triggers convection. With the box model
they argue that the oscillation occurs when the time scale 
for the surface freshening
is shorter than the time scale of the subsurface warming. 
However, because the advection of fresh water 
is described by a simple 
relaxation condition, it is not clear where the ocean 
dynamics come in. The results of the box model suggests that 
the suppression of convection could be caused by a 
surface freshening solely due to the increased 
salinity gradient. Our experiments with the ocean model
with a stationary circulation proof that this is not the case. 
Furthermore, with this mechanism salt perturbations 
would result in a phase shift of the oscillation. 
If, for example, one triggers convection before 
the subsurface heat flux actually destabilizes the 
stratification (between phase 2 and 
phase 3 in Fig. \ref{h3.yin.mech}) the system would immediately
switch to the convective phase (phase 1). 
The perturbed run would therefore be out of phase with the 
run that is not perturbed. Such a phase shift was however not 
observed in the perturbation experiments with the ocean model.
In contrast, even with fairly large perturbations we were
not able to cause a significant phase shift.
Yin's box model will be analyzed in the appendix. It will 
be shown that oscillations in a box model only occur in an
unlikely part of the parameter space. In a more realistic 
parameter space, convection is selfsustaining and no 
oscillations occur. Ocean dynamics are therefore important.

To conclude, all our experiments indicate
that ocean dynamics play an essential role. 
This has an important impact for the time scale 
of the oscillation. In Yin's case the time scale
is mainly determined by the rate of subsurface warming.
But, since the warming of the subsurface is a relatively 
fast process (order of a few years), it cannot explain the 
(inter)decadal time scale of the oscillation.  
In our view, this time scale is mainly determined by the 
ocean dynamics. It is mainly determined by the time it takes for the surface 
circulation to responds to anomalous convective activity.
  
\section{Summary and Conclusions \label{h3.sec.conclusions}}

We investigated the mechanism causing a decadal oscillation in an ocean model
forced by MBCs. The oscillation is similar to the oscillation described in 
WS91, Wea93 and also in YS95. The oscillation is characterized 
by large fluctuations in convective activity and atmospheric heatflux in a relatively 
small area in the northeastern part of the basin. The meridional overturning and the 
meridional heattransport show significant variations on a decadal
time scale in response to the convective activity. 
The phase relationship between convective activity and the strength of
meridional overturning (given by the amplitude of the positive overturning cell with water flowing
poleward at the surface) shows a significant lag. The maximum of the overturning
coincides with a minimum in convective activity. An overturning cell at the polar
boundary with water flowing equatorward at the surface also displays large 
fluctuations. This negative cell is strongest
just before the deep convection at the eastern boundary disappears. The cell
disappears just after the deep convection is turned off. 
 
The mechanism can be summarized as follows. At the end of the convective phase
an increasing amount of fresh water is flowing from the polar
boundary southeastward into the convective area. The southward flow is shown in 
the meridional overturning by the polar cell. Eventually, convection is 
completely shut off. The propagation as a Kelvin wave of the main area of downwelling 
at the eastern boundary  causes a rapid collapse of the negative 
cell. The southeastern flow at the surface weakens, and the advection 
of fresh water decreases. The resulting surface salinity rise, together with 
a rapid subsurface warming, initialize convection at the eastern boundary. Thereafter
convection rapidly expands because of a convective instability. Convection
mixes heat and salt to the surface; the heat is rapidly released to the 
atmosphere whereas the surface salinity anomaly remains, expands horizontally and 
triggers convection in the neighboring gridcells. The positive surface salinity 
anomaly due to convection causes an increase in the surface salinity gradient. The increased
surface salinity gradient, together with increasing southeastward velocities 
at the surface, cause an increase in the net advection of fresh water into the convective 
area. In response to the slowly increasing influx of fresh water and a rapid cooling of the 
subsurface the convective activity again decreases.

With this paper we aimed to answer the question as to whether the ocean dynamics 
play an essential role in the oscillation. With a conceptual model we 
showed that a phase shift between the influx of fresh water 
into the convective area and the convective activity causes oscillatory behavior. 
The phase lag between the advection and the convective 
activity is introduced by the dynamic response of the surface 
velocity. The negative polar cell increases in strength during the convective phase, 
and therefore advects more fresh water into the convective area, which reduces the 
convective activity.
 
Our results strongly suggests that the period
of the oscillation is determined by the the time scale at which the advection of fresh water 
into the convective area responds to changes in convective activity. The rate 
of the subsurface warming has only a small impact on the time scale of the oscillation.
In Lenderink (1996) we explored this point further. It furthermore turns out that the 
amplitude of the oscillation is mainly determined by the rate of subsurface warming.
This contradicts the results of Yin (1995) and YS95 who argued that the 
time scale of the oscillation is mainly set by the subsurface warming. In Yin (1995)
the subsurface warming actually destabilizes the stratification. In our case 
a warming of the subsurface only is a necessary, but not sufficient condition
to trigger of convection.

The main question is therefore what causes the change of the flow into the convective 
area. We first considered the convective phase. Associated with the
convective activity is upwelling and downwelling at the 
eastern and northern boundary, which introduces a positive density anomaly 
north of the convective area and a negative density anomaly south of the 
convective area (see also Moore and Reason 1993). 
Horizontal advection also displaces the density anomaly northward. 
The net results of these processes is that - in response to anomalous convective 
activity - the densest water tends to be trapped 
in the northeastern corner of the basin. The anomalous surface flow associated with this 
density anomaly is now a southeastward flow into the convective area which acts to suppress 
convection. During the nonconvective phase the occurrence of a Kelvin
wave plays an important role. It re-initialiazes convection and it resets the 
southeastern surface flow.                  
                                                                                      
Lateral boundaries determine the dynamic response
of the ocean model to a large extent. Because the processes
at the boundary are obviously not well described in   
coarse resolution ocean models, this fact causes
our major concern. The model bahavior could 
change dramaticaly when the processes at the boundary 
are represented more realistically. 
In a recent paper, Winton (1996b) proposes that  
decadal oscillations are greatly damped when bottom 
topography is included. The inclusion of bottom topography
changes the dynamic response because potential energy 
can be directly converted into the barotropic mode with bottom
topography, whereas without bottom topography potential
energy can only be converted into the baroclinic mode. 
The adjustment of the thermal winds to the no-normal
flow condition at the eastern boundary changes
with the inclusion of bottom topography. 
This agrees with the results of Moore and Reason (1993). 
They found a decadal oscillation in a global ocean model
with a flat bottom, but did not find an oscillation with realistic
bottom topography. 

Zhang et al. (1995) found an oscillation in an ocean model coupled to a thermodynamical
ice model. They contributed the oscillation to the thermal insulating effect of 
sea-ice. However, their oscillation
shows the same phase relationship between the convective activity and the surface
flow. In addition, the surface salinity also shows a sharp gradient between the convective
area and the northwestern boundary. 
This suggest that the mechanism described in this paper might operate.
Therefore, we also performed a few experiments with an ice model included.  
The results (not described in this paper) seem to confirm our hypothesis.  
Oscillations caused by the mechanism described in this paper 
with a time scale of 15-20 years were obtained with a freshwater forcing that 
causes a pool of fresh water in the northwestern
corner. It even seems that oscillations are more easily invoked 
because the ice-layer pushes convection south of the polar boundary, which 
facilitates the formation of the negative overturning cell at the polar boundary. 
These results suggest that the mechanism described in this paper is fairly 
robust in coarse resolution ocean models. 
It remains however to be veryfied whether the oscillation also 
occurs when the processes occuring at the boundary are represented more 
realistically.  \\ \\

\addcontentsline{toc}{section}{Appendix A}
\markright{\small{\sc Appendix A}}
{\large \bf \begin{center} Appendix A \\ \end{center}}

\section*{A two box model \label{h3.sec.box}}

The oscillation that is found by Welander (1982) in a two box model has often been 
used to explain the occurrence of decadal oscillations in ocean models.
The oscillation occurs when, with a certain heat and freshwater forcing,
both the convective and the nonconvective equilibrium 
state do not exist. This results in
oscillations between the convective and the nonconvective state.  

In a box model the net effect of the advective and diffusive exchange 
between the boxes and the surrounding ocean has to be parameterized. 
A relaxation condition or a fixed flux are often used as simple 
parameterizations of these fluxes. 
Using a relaxation condition for the advection of salt into the surface
box, only the static response of 
advection to changes in convective activity is represented.
We will show that in such a box model, using a forcing that is realistic 
in the sense that it forces an halocline and inverse thermocline structure,
the occurrence of oscillations is unlikely.

\subsection*{The box model}

In this section we solve a simple two box model. For simplicity we assumed 
that the volume of the surface equals that of the subsurface box.
This assumption does not affect our conclusions as long as the volumes
of both boxes are of the same order.
Both the temperature (T) and salinity (S) of the surface 
(labeled with a subscript u) and the subsurface  
(labeled with a subscript l) box are dynamical variables. 
The equations are given by:
\begin{eqnarray}
\frac{dT_u}{dt} & = & \alpha(T_{atm}-T_u) + \tau \Delta (T_l-T_u) \nonumber \\
\frac{dT_l}{dt} & = & q_l^T (T_l^b-T_l) + \tau \Delta (T_u-T_l)     \nonumber \\
\frac{dS_u}{dt} & = & q_u^S (S_u^b-S_u) + \tau \Delta (S_l-S_u)     \nonumber \\
\frac{dS_l}{dt} & = & q_l^S (S_l^b-S_l) + \tau \Delta (S_u-S_l) \label{h3.eq.box}
\end{eqnarray}
The notation is similar to the one used in LH94. The first terms on the right hand side
denote the effect of advection, diffusion and atmospheric forcing.
The constants $T_l^b$, $S_u^b$ and $S_l^b$ can be considered to be the characteristics of the water
flowing into the boxes. The advective timescale $q^{-1}$ is determined by flow velocity. 
Here, the superscript denote either salinity (S) or temperature (T). 
The surface temperature $T_u$ is relaxed with a restoring constant $\alpha$ to the 
atmospheric temperature $T^{atm}$. The second terms on the right hand side represent
convective mixing, which occurs in case of an unstable stratification. 
The convective timescale is given by $\tau^{-1}$, and $\Delta$ is either 0 or 1 
dependent on the stratification. The equation of state is linear:
\[ \rho=\rho_o - k^T T + k^S S, \]
where $\rho_o$, $k^T$ and $k^S$ are positive constants.

\subsection*{Solutions of the box model}

\begin{figure}[t]
\centerline{\psfig{figure=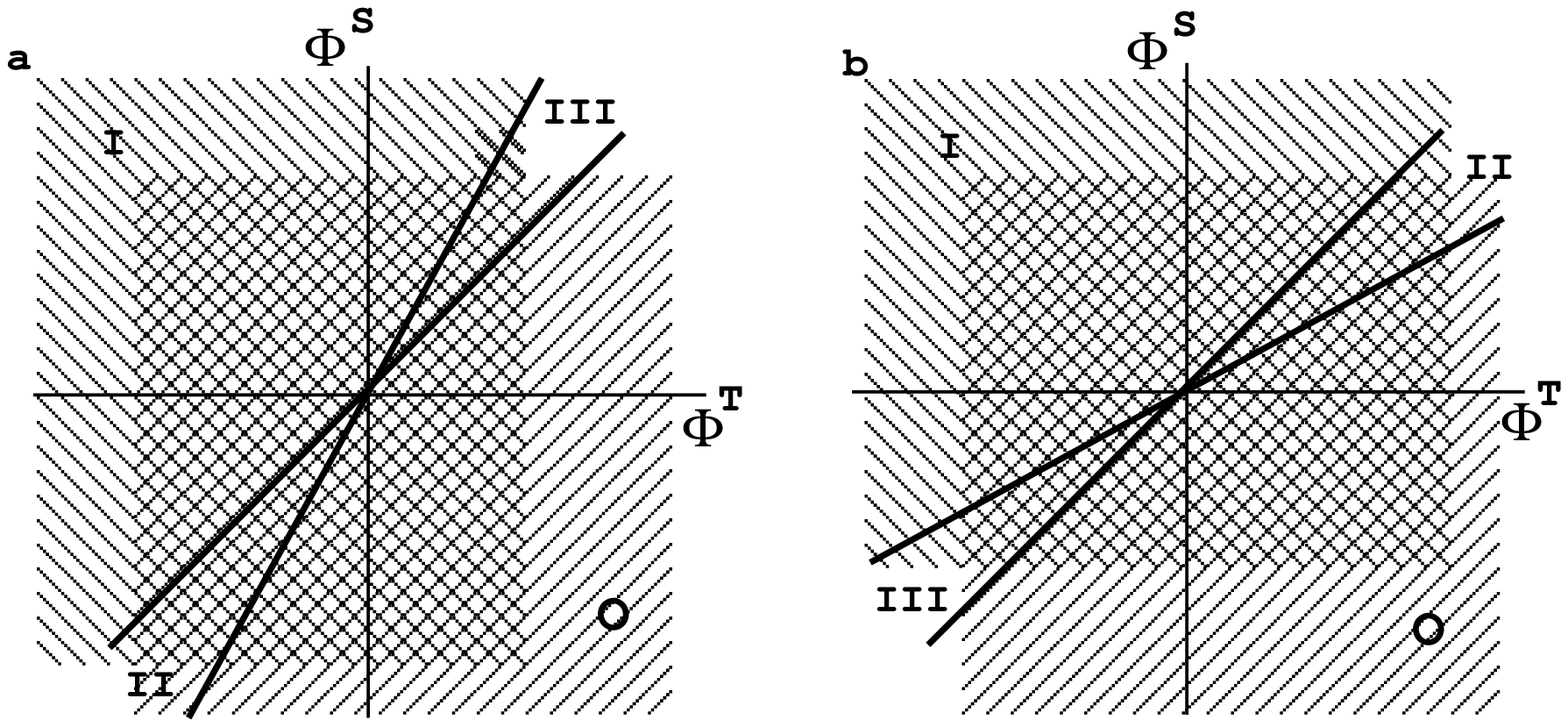,width=10cm}}
\vspace{0.5cm} 
\center{\parbox{14cm}{\renewcommand{\baselinestretch}{1.1}
\caption{\label{h3.wea.boxsol}\small 
Solutions of the box model with the nonconvective regime O, the convective
regime I, the regime with both the convective and the nonconvective
solution II, and the oscillatory regime III 
}}}
\end{figure}

We solve these equations similar to the method used in LH94. 
The stationary solutions are determined by putting the lefthand sides  of (\ref{h3.eq.box}) to zero.
Then we solve the system for the nonconvective equilibrium ($\Delta=0$) and the convective
equilibrium ($\Delta=1$). For the nonconvective equilibrium this yields:
\[ T_l-T_u = T_l^b - T_{atm} \]
\begin{equation} S_l-S_u = S_l^b - S_u^b, \label{h3.eq.stat0}  \end{equation}
and for the convective equilibrium:
\[ T_l-T_u = \left[1 + \frac{\tau}{\alpha} + \frac{\tau}{q_l^T}\right]^{-1} (T_l^b - T_{atm}) \]
\begin{equation}
  S_l-S_u = \left[1 + \frac{\tau}{q_u^S} + \frac{\tau}{q_l^S}\right]^{-1} (S_l^b - S_u^b).
                                      \label{h3.eq.stat1}  \end{equation}
The nonconvective equilibrium now exists if and only if:
\[ k_T (T_l^b-T_{atm})  < k_S ( S_l^b - S_u^b ) \]
We abbreviate this to 
\begin{equation} \Phi^T < \Phi^S. \label{h3.eq.condnon} \end{equation}
Similarly, the convective equilibrium exists if and only if:
\begin{equation}  
\left[1 + \frac{\tau}{\alpha} + \frac{\tau}{q_l^T}\right]^{-1} \Phi^T 
         > \left[1 + \frac{\tau}{q_u^S} + \frac{\tau}{q_l^S}\right]^{-1} \Phi^S . \label{h3.eq.condcon} 
\end{equation}
If we consider the space spanned by $\Phi^T$ and $\Phi^S$ the possible solutions 
can be visualized as follows. The equations \ref{h3.eq.condnon} and \ref{h3.eq.condcon}
divide this space into four parts. 
If we impose the following condition on the timescales: 
\begin{equation}
             \frac{\tau}{\alpha} + \frac{\tau}{q_l^T} 
         >   \frac{\tau}{q_u^S} + \frac{\tau}{q_l^S}, \label{h3.eq.conosc}
\end{equation} 
the four parts are as shown in Fig. \ref{h3.wea.boxsol}a. 
If this condition is not fulfilled the four parts are as shown in Fig. \ref{h3.wea.boxsol}b.
Similar to LH94 we labeled these four parts with regime 0, I, II and III. In regime 0
only the nonconvective state exists; in regime I only the convective state exists;
in regime II both the convective and the nonconvective state exists, and in regime
III no equilibrium solutions are found and the system oscillates.
It should be noted that, strictly speaking, the condition given by Eq. \ref{h3.eq.conosc}
is only determined by the restoring timescales valid for the convective phase. 

\subsection*{Convective oscillations in a box model}

The Welander oscillation is similar to the oscillations we obtain in regime III.
We only consider positive values for $\Phi^T$ and $\Phi^S$ 
because only in this part of the parameter space is a 
halocline and an inverse thermocline established.
If the decadal oscillations found in the ocean model 
are identified with regime III, the condition given by Eq. \ref{h3.eq.conosc} 
therefore has to satisfied.

The restoring constants for the subsurface box  $q_l^S$ and $q_l^T$ are mainly determined
by the flow velocity in the ocean. There is a priori no reason to assume these 
constants to be different. If we therefore assume  $q_l^S = q_l^T$ Eq. \ref{h3.eq.conosc} 
becomes:
\[ \alpha < q_u^S. \]
Because the restoring to the atmospheric temperature is fast compared to the advective 
timescale this conditions is not fulfilled. 

In the convective phase the vertical temperature and salinity gradient are determined
by the timescale of the advective (atmospheric) processes and the time scale of the 
convective process. 
As show by Eq. \ref{h3.eq.stat1} the vertical gradient increases with increasing advective 
timescales and a decreasing convective timescale. The longest advective timescale
mainly determines the vertical gradient. The temperature gradient is mainly
determined by $q^T_l$ because the atmospheric restoring is fast.
Because $q^S_u$ and $q^S_l$ are both advective timescales and are thus of the same
order, the salinity gradient is determined by both $q^S_u$ and $q^S_l$.
It is unlikely that the salinity gradient stabilizes the stratification in the convective state
as Eq. \ref{h3.eq.conosc} is not likely to be fulfilled. 
The parameter space is given by Fig. \ref{h3.wea.boxsol}b.  
Oscillations as a results of regime III 
do not occur with a forcing that establishes a halocline and an inverse thermocline. 
Instead the system is in that part of the parameter space where regime II
exists. Convection is selfsustaining and hysteresis occurs.

One could therefore conclude that either the formulation of the forcing
for the box model is not appropriate, or an external force is missing to 
obtain convective oscillations. One could for example assume that 
the characteristics $T_l^b, S_l^b$, and $S_u^b$ of the water entering the area considered depend
on the convective activity because the direction of the flow changes due to the 
changes in the convective activity.  
In the ocean model this is visible at the northeastern boundary 
between t = 4 and t = 6 years, when convection is completely suppressed. At t = 4 years 
a strong southward flow transport relatively fresh water into the convective area,
which suppresses convection completely just after t = 5 years. When the velocity field 
does not change, convection would remain off, even in spite of the subsurface warming
when convection is off. The nonconvective state is selfsustaining.
Due to the boundary Kelvin wave, however, the surface flow reverses sign, and the northward
flow advects much saltier water into the area. This rapidly re-initializes convection. 
The subsurface warming is therefore a necessary, but not sufficient condition
to give a convective oscillation.
  
\subsection*{Yin's oscillation}

In Yin (1995) a similar box model is solved. In contrast to our analysis
Yin obtained decadal oscillations. 
Yin however assumed that the subsurface salinity is fixed. The salinity 
gradient when convection occurs is now determined by the rate of surface 
freshening only, whereas without this assumption it is also limited by the 
advective timescale for the subsurface. Mathematically, it
is equivalent with using an infinite restoring constant for the subsurface salinity. 
Using this and assuming that 
$\alpha >> q_l^T$, Eq. \ref{h3.eq.conosc} becomes
\[ q_l^T < q_u^S. \]
For positive $\Phi^T$ and $\Phi^S$ regime III therefore only occurs if the timescale
of the surface freshening is shorter than the timescale of the subsurface warming. 
This result is in accordance with Yin (1995). \\ \\ \\

{\em Acknowledgments}. This work was initiated during a visit
of the first author to Andrew Weaver and his group. 
Many thanks. Thanks are also due to Nanne Weber, Wim Verkley and Frank Selten 
for comments on earlier versions of the manuscript. 
This work has been supported by the Netherlands 
Organization for Scientific Research (NWO) project VVA.

\pagebreak

\pagebreak

\noindent {\Large Figures} \\ \\

\clearpage

\begin{thebibliography}{99}

\bibitem{Bjerknes}
Bjerknes, J., 1964: 
Atlantic air-sea interaction. 
{\em Advances in Geophysics}, Academic Press, 1-82.

\bibitem{Broecker}
Broecker, W. S., D. M. Peteet, and D.  Rind, 1985: 
Does the ocean-atmosphere system have more than one stable mode of operation? 
{\em Nature}, {\bf 315}, 21-26.

\bibitem{Bryan}
Bryan, K., 1984: 
Accelerating the convergence to equilibrium of ocean-climate models. 
{\em J. Phys. Oceanogr}., {\bf 14}, 666-673.

\bibitem{Cai}
Cai, W., 1995: 
Interdecadal variability driven by mismatch between surface flux 
forcing and oceanic freshwater/heat transport.
{\em J. Phys. Oceanogr}., {\bf 25}, 2643-2666.

\bibitem{Cai2}
Cai, W., and S. J. Godfrey, 1995: 
Surface Heatflux parameterizations and the variability of the 
thermohaline circulation. 
{\em J. Geophys. Res.}, Vol. 100, No. C6, 10,679-10,692.

\bibitem{Cai3}
Cai, W., R. J. Greatbatch, and S. Zhang, 1995: 
Interdecadal variability in an ocean model driven by a small, 
zonal redistribution of the surface buoyancy flux. 
 {\em J. Phys. Oceanogr}., Vol. 25, No. 9, 1998-2010.

\bibitem{Chen}
Chen, F., and M. Ghil., 1995: 
Interdecadal variability of the thermohaline circulation and high-latitude 
surface fluxes. 
{\em J. Phys. Oceanogr.},  Vol. 25, No. 11, 2547-2568.

\bibitem{Delworth}
Delworth, T., S. Manabe, and R. J. Stouffer, 1993:
Interdecadel variability of the thermohaline circulation in a coupled
ocean-atmosphere model.
{\em J. Climate}, {\bf 6}, 1993-2011.

\bibitem{Drijfhout}
Drijfhout, S., C. Heinze, M. Latif, and E. Maier-Reimer., 1996:
Mean Circulation and Internal Variability in an Ocean Primitive Equation Model.
{\em J. Phys. Oceanogr}., Vol. 26, No. 4., 559-580. 
  
\bibitem{Greatbatch1}
Greatbatch, R. J., and S. Zhang, 1995: 
An interdecadal oscillation in an idealized ocean basin forced by constant heat flux. 
{\em J. Climate}, Vol. 8. No. 1, 81-91. 


\bibitem{Greatbatch2}
Greatbatch, R. J., and K. A. Peterson, 1995: 
Interdecadal variability and ocean thermohaline adjustment. 
submitted to J. Geophys. Res.

\bibitem{Haney}
Haney, R. L., 1971: 
Surface thermal boundary condition for ocean circulation models. 
{\em J. Phys. Oceanogr}., {\bf 1}, 241-248.

\bibitem{Huang}
Huang R. X., and L. Chou, 1994: 
Parameter sensitivity study of the saline circulation. 
{\em Clim. Dyn}., {\bf 9}, 391-409.

\bibitem{Huang}
Huang R. X., 1993: 
Real freshwater flux as a natural boundary conidtion for the 
salinity balance and thermohaline circulation forced by evaporation nad precipitation. 
{\em J. Phys. Oceanogr}., {\bf 23}, 2428-2446. 

\bibitem{Kushnir}
Kushnir, Y., 1994:
Interdecadal variations in the North Atlantic sea surface temperature 
and associated atmospheric conditions. 
{\em J. Climate}, Vol. 7, 141-157.

\bibitem{LH}
Lenderink, G., and R.J. Haarsma, 1994: 
Variability and multiple equilibria of the thermohaline circulation 
associated with deep water formation.
{\em J. Phys. Oceanogr}., Vol. 24, No. 7, 1480-1493.

\bibitem{LH2}
Lenderink, G., and R.J. Haarsma, 1996: 
Modeling Convective Transitions in the presence of sea-ice. 
{\em J. of Phys. Oceanogr}., Vol 26, {\bf 8}, p 1448-1467.

\bibitem{LH3}
Lenderink, G., 1996:
Decadal oscillations in a box model with a dynamical feedback.
to be submitted to J. Phys. Oceanogr.

\bibitem{myers}
Myers, P. G. and A. J. Weaver, 1992: 
Low-frequence internal oceanic variability under seasonal forcing. 
{\em J. Geophys. Res}., {\bf 97}, 9541-9563.

\bibitem{moore}
Moore, A. M., and J. C. Reason, 1993: 
The response of a global ocean general circulation model to climatological 
surface boundary conditions for temperature and salinity. 
{\em J. Phys. Oceanogr}., Vol. 23, No. 2, 300-328.


\bibitem{Power3}
Power, S. B., and R. Kleeman, 1994: 
Surface heat flux parameterization and the response of OGCMs to high latitude freshening. 
{\em Tellus}, {\bf 46A}, 86-95.


\bibitem{rahm}
Rahmstorf, S., 1994: 
Rapid climate transitions in a coupled ocean atmosphere model. 
{\em Nature}, Vol. 372, 82-85.

 
\bibitem{Weaver}
Weaver, A. J., and E. S. Sarachik, 1991 (WS91): 
Evidence for decadal variability in an ocean general circulation model: 
an advective mechanism. 
{\em Atmosphere-Ocean}, {\bf 29}, 197-231.

\bibitem{Weaver2}
Weaver, A.J., J. Marotzke, P. F. Cummins, and E. S. Sarachik, 1993 (Wea93):
Stability and variability of the thermohaline circulation.
{\em J. Phys. Oceanogr}., {\bf 1}, 39-60.

\bibitem{Welander}
Welander, P., 1982: 
A simple heat salt oscillator.
{\em Dyn. Atmos. Oceans}, {\bf 6}, 233-242.

\bibitem{weisse}
Weisse, R., U. Mikolajewicz, and E. Maier-Reimer, 1994: 
Decadal variability of the North-Atlantic in an ocean general circulation model. 
{\em J. Geophys. Res}., {\bf 99}, 12411-12421.

\bibitem{Winton}
Winton, M., 1996a: 
On the role of horizontal boundaries in parameter sensitivity 
and decadal-scale variability of coarse resolution ocean general circulation models. 
{\em J. Phys. Oceanogr}., Vol. 26, No. 3, 289-304.

\bibitem{Winton2}
Winton, M., 1996b: 
The damping effect of bottom topography on internal decadal scale 
oscillations of the thermohaline circulation.
{\em J. Phys. Oceanogr}., Submitted.

\bibitem{yang}
Yang, J., and J. D. Neelin, 1993: 
Sea ice interaction with the thermohaline circulation. 
{\em Geophys. Res. Lett}., {\bf 20}, 217-220.

\bibitem{YinS}
Yin. F., and E. S. Sarachik, 1995: 
Interdecadal thermohaline oscillations in a sector ocean general circulation model: 
advective and convective processes. 
{\em J. Phys. Oceanogr}., {\bf 25}, 2465-2484.

\bibitem{YinS}
Yin. F., 1995: 
A mechanistic model of the ocean interdecadal thermohaline oscillations. 
{\em J. Phys. Oceanogr}., {\bf 25}, 3239-3246.

\bibitem{Zhang}
Zhang, S., R. J. Greatbatch, and C. A. Lin, 1993: 
A reexamination of the polar halocline catastrophy and implications 
for coupled ocean-atmosphere modeling. 
{\em J. Phys. Oceanogr}., {\bf 2}, 287-299.

\bibitem{Zhang2}
Zhang, S., C. A. Lin, and R. J. Greatbatch, 1995:
A decadal oscillation due to the coupling between an ocean model an a 
thermodynamical sea-ice model. 
{\em J. Mar. Res}, {\bf 53}, 79-106. 


\end{thebibliography}
\end{document}